\documentstyle[preprint,aps,eqsecnum,epsfig]{revtex} 
\newcommand{\be}{\begin{equation}}
\newcommand{\ee}{\end{equation}}
\newcommand{\bea}{\begin{eqnarray}}
\newcommand{\eea}{\end{eqnarray}}
\newcommand{\dbarp}{\frac{d^3p}{(2\pi)^3}}
\newcommand{\dbarq}{\frac{d^3q}{(2\pi)^3}}
\tighten 
\begin{document} 
\preprint{PITT-00; LPTHE/00-16} 
%\draft  
\title{\bf RELAXING NEAR THE CRITICAL POINT}   
\author{\bf D. Boyanovsky$^{(a,b)}$, H. J. de Vega$^{(b,a)}$,
 M. Simionato$^{(b,c)}$}
\address
%\affiliation
{(a) Department of Physics and Astronomy, University of 
Pittsburgh, Pittsburgh  PA. 15260, U.S.A\\(b) LPTHE, Universit\'e
Pierre et Marie Curie (Paris VI) et Denis Diderot  
(Paris VII), Tour 16, 1er. \'etage, 4, Place Jussieu, 75252 Paris
cedex 05,France\\(c) 
 INFN, Gruppo Collegato di Parma, Italy}
\date{\today}
\maketitle 
\begin{abstract} 
Critical slowing down of the
relaxation of the order parameter has phenomenological consequences in
early universe cosmology and in ultrarelativistic heavy ion
collisions. We study the relaxation rate of long-wavelength fluctuations of the
order parameter in an $O(N)$ scalar theory near the critical point to model  
the non-equilibrium dynamics of critical fluctuations near  the chiral
phase transition. A lowest order perturbative calculation (two loops
in the coupling $\lambda$) reveals the breakdown of perturbation
theory for long-wavelength fluctuations in the critical region and the
emergence of a hierarchy of scales with hard $q\geq T$, semisoft $T\gg q \gg
\lambda T$ and soft $\lambda T\gg q$ loop momenta which are widely separated
in the weak coupling limit. A non-perturbative resummation is
implemented to leading order in the large $ N $ limit which reveals
the infrared renormalization of the static scattering amplitude and
the crossover to an effective three dimensional theory for  
the soft loop momenta near the critical point. The effective three
dimensional coupling is driven to the Wilson-Fisher three dimensional
fixed point in the soft limit. This resummation provides an infrared
screening and for critical fluctuations of the order parameter with
wave-vectors $ \lambda T\gg k \gg k_{us} $ or near the critical temperature 
$\lambda T \gg m_T \gg k_{us}$ with the {\bf ultrasoft} scale $k_{us}
= \frac{\lambda T}{4\pi}\exp{[-\frac{4\pi}{\lambda}]}$ the relaxation  
rate is dominated by {\em classical} semisoft loop momentum leading to
$ \Gamma(k,T)= \lambda T/(2\pi N) $. For wavectors $ k\ll k_{us} $ the
damping rate is dominated by hard loop momenta and given by
$ \Gamma(k,T)= 4\pi T/\left[3N \ln(T/k)\right] $. Analogously, for
homogeneous fluctuations in the ultracritical region $ m_T \ll k_{us} $ 
the damping rate is given by $  \Gamma_0(m_T,T)=4\pi T/\left[3N
\ln(T/m_T)\right] $. Thus critical slowing down emerges
for ultrasoft fluctuations. In such regime the rate is independent of
the coupling $\lambda$ and both perturbation theory
and the classical approximation within the large $N$ limit break down.  
The strong coupling regime and the shortcomings of the quasiparticle
interpretation are discussed.   
\end{abstract} 
\pacs{11.10.Wx;12.38.Mh;11.15.Pg;64.60.Ht}
%\maketitle 
\section{Introduction and Motivation} 

The program of relativistic heavy ion collisions both at Brookhaven
and at Cern seeks to understand the phase diagram of QCD in conditions
of temperatures that were achieved during the first 10$\mu$s after the
Big Bang or densities several times 
that of nuclear matter which could exist at the center of neutron stars. Current theoretical understanding\cite{QCD} leads to the conclusion that QCD could undergo two phase transitions: a confinement-deconfinement (or hadronization)
and the  chiral phase transition. Current lattice data seem to suggest
that both occur at about the same 
temperature $T_c \approx 160 \mbox{MeV}$\cite{QCD}. The consensus
emerging in the field is that several types of observables will have
to be studied simultaneously and event-by-event analysis of data will
have to be carried out to extract unambiguous signals both hadronic
and electromagnetic to reveal the presence of a Quark-Gluon Plasma
phase. Recent results reported from CERN-SPS\cite{cern}  seem to
indicate a strong evidence for the existence of the  QGP in Pb-Pb
collisions, and 
RHIC at Brookhaven will begin operation soon with Au-Au collisions
with four dedicated detectors capable of event-by-event 
analysis of hadronic and electromagnetic observables.  

For QCD with only two flavors of massless quarks (u,d) it has been
argued\cite{o4pis,o4raja} that the chiral phase transition at finite
temperature but vanishing baryon number density is of second order and
described by the universality class of O(4) Heisenberg
ferromagnets. It has also been suggested recently that at finite
baryon density there is a 
second order critical point described by the Ising universality
class\cite{ising}. Second order critical points are characterized by
strong critical long-wavelength fluctuations and a diverging
correlation length that  could lead to 
important experimental signatures\cite{QCDPT}. These signatures would
be akin to  critical opalescence near  
the critical point in binary fluids\cite{QCDPT} and could be observed
in an event-by-event analysis of the fluctuations of the 
charged particle transverse momentum distribution (mainly pions)\cite{QCDPT}. 
These fluctuations are characterized by the typical correlation length
of the order parameter and it has been  suggested that the phenomenon
of critical slowing down, 
ubiquitous near the critical point of second order phase transitions,
can lead to strong departures from equilibrium that 
will determine the value of the correlation length when
long-wavelength fluctuations freeze out\cite{rajaslow}. Critical 
slowing down of long-wavelength fluctuations near a second order
critical point is the statement that the 
long-wavelength Fourier components of the order parameter relax very
slowly towards equilibrium\cite{MA}. In mean-field 
theory in {\em classical critical phenomena},  the relaxation time diverges proportional
 to the susceptibility
near the critical temperature but thermal fluctuations renormalize the
relaxation time to be of the form $\tau(\vec k=0) \propto \xi^z$ with  $\xi$ the
correlation length, or at critical point $\tau(k) \propto k^z$ with $z$ a 
dynamical critical exponent  \cite{MA,hoh}. Another similar manifestation of an
anomalously slow relaxation of long-wavelength fluctuations arises in
weakly first order phase transitions when the system enters into the
mixed phase where the (isothermal) speed of sound, which determines
the velocity of propagation of long-wavelength pressure waves, becomes
anomalously small resulting in the softest point of the equation of state. 
 In this case there have also been suggestions that there are experimental
consequences of this softening in relativistic heavy ion collisions in
observables  related to collective flow and the transverse momentum
distributions of particles at freezeout\cite{shuryak}.  

In classical normal fluids near the critical point  the vanishing of
the (isothermal) speed of sound, critical  
opalescence (strong scattering of light by long-wavelength
fluctuations) and critical slowing down are all related\cite{forster},
and in 
ferromagnets  the spin diffusion constant vanishes near the critical
point again signaling critical slowing down\cite{forster}.  

The softening of the equation of state near the critical point of QCD
 could also have important cosmological implications. When the 
 the QGP enters the mixed phase with hadrons, the speed  of sound
 becomes anomalously small and the time scale for propagation of
 pressure waves over a given critical wavelength becomes longer  
than the free fall time for gravitational collapse 
which is then unhindered by the pressure of the hadronic
gas. This could lead to
 the formation of primordial black holes\cite{jedamzik} with 
a possible imprint in the acoustic peaks in the cosmic microwave
 background\cite{widerin}. Other possible cosmological relics from 
the QCD phase transition with a mixed phase had been predicted, from
 strange quark nuggets to MACHO's\cite{raja,mau}.  

A familiar argument is typically invoked to state that while the QCD
phase transition in the Early Universe occurred in local thermodynamic
equilibrium (LTE) this {\em may} not be the case in Relativistic Heavy
Ion Collisions. The argument compares the typical collisional
relaxation 
time scale obtained from a  strong interaction process 
$\tau_{coll} \sim 10^{-22}~\mbox{secs}$ to the time scale for cooling
near the critical temperature $\sim 160 \mbox{MeV}$, i.e. ${T/
\dot{T}} \sim H^{-1} \sim 10^{-5}~\mbox{secs}$.
 The argument is that since $\tau_{coll}
\ll   H^{-1}$ the phase transition occurs in LTE
in cosmology, whereas in relativistic heavy ion collisions at RHIC and LHC
energies these time scales will be comparable. However this argument
completely neglects the possibility that long-wavelength fluctuations
could undergo 
critical slowing down and freeze out, i.e. fall out of local thermal
equilibrium, even {\em before} the phase transition. The freeze-out of long-wavelength fluctuations
during the phase transition could  result in 
important non-equilibrium effects on the size and distribution of
primordial black holes or any other cosmological relic just as they
could lead to 
important observables in the momentum distributions of charged pions
in relativistic heavy ion collisions\cite{QCDPT,rajaslow,shuryak}.  

Indeed there are simpler experimental situations where this is the
case, in typical normal fluids a collisional relaxation time (away 
from the critical point) is of the order of $10^{-9}~\mbox{secs}$
while near the critical point (even at $10\%$ of the critical
temperature) critical slowing down becomes very dramatic and
thermalization time scales become of the order of minutes if not
hours\cite{forster,stanley,boyleshouches}. 

Thus the phenomenological importance of critical slowing down for the
QCD phase transition both in Relativistic Heavy Ion Collisions as well
as in Early Universe Cosmology motivates us to study this phenomenon
in a model Quantum Field Theory that bears on the low energy (chiral)
phenomenology of QCD, the $O(N)$ linear sigma model. Furthermore the
study of critical 
slowing down is the precursor to a more complete program to understand
transport phenomena  and the relaxation of hydrodynamic modes at or near a critical point.    

{\bf Goal:} Our goal is to provide a consistent microscopic
description of critical slowing down at or near criticality directly from
an underlying quantum field theory that is at least phenomenologically
motivated to study the 
QCD phase transitions. This will be a first step in a program that
seeks to offer a consistent description of transport near 
critical points that eventually may be merged with a hydrodynamic
description to obtain a more reliable picture of critical 
phenomena near the deconfinement and chiral phase transitions and an
assessment of the potential phenomenological observables both in early universe
cosmology as well as in relativistic heavy ion collisions.  We begin this program in this
article by focusing on the relaxation rate of long-wavelength fluctuations of the order 
parameter  at and near  the critical point in a consistent non-perturbative framework. 

{\bf Strategy:} We begin our study of critical slowing down by
analyzing the relaxation rate of long-wavelength fluctuations of the
order parameter at and near the critical point in  an $O(N)$ scalar field
theory, which is a phenomenological arena to study the 
relaxation of sigma mesons and pions. Our first step is to obtain the
relaxation rate to lowest order in perturbation theory 
(two loops). This calculation reveals clearly the breakdown of the
perturbative expansion for long-wavelength fluctuations at or near the critical point as a result of
the strong infrared behavior for soft loop momentum and the necessity
for a non-perturbative treatment. We then implement a non-perturbative
resummation of bubble-type diagrams via the large $ N $ approximation
to obtain the damping rate in the next-to-leading order in the large $ N $
limit. The resummation implied by the large $ N $ limit to order $1/N$
is akin to that obtained via the renormalization group with the 
one loop beta function and reveals the softening of the scattering
amplitude and the  crossover to an effective three dimensional theory
for momenta $q   \ll  \lambda T$ with $\lambda$ the quartic coupling.   

{\bf Summary of main results:}  

We have obtained the relaxation rate for long-wavelength fluctuations of the
order parameter at the critical point and for homogeneous fluctuations
near criticality both to lowest order in perturbation theory (two loops)  
and near the critical point  to next to leading order in the large $N$ limit.  

The two-loop results for the relaxation rate for a fluctuation of
wavevector $\vec k$ of the order parameter at the critical point is
found to be $\Gamma(k,T) \propto \lambda^2 T^2/k$ whereas near the
critical point, homogeneous fluctuations (with $\vec k=0$) relax with a
rate $ \Gamma_0(m_T,T) \propto \lambda^2 T^2/m_T $. Here $ m_T \propto
|T-T_c|^{1/2}\ll T_c$ is the effective thermal mass. These results
clearly reveal the breakdown of the perturbative expansion in the long
wavelength limit $ k \rightarrow 0 $ at $ T = T_c $ and for $ T \to
T_c $ and $ k = 0 $. 

A detailed analysis of the different
contributions to these results for the relaxation rate shows that the rate is 
dominated by very soft  loop momentum $q   \ll  \lambda T$ which in
the weak coupling limit $\lambda    \ll  1$ are classical. The
implementation of a non-perturbative resummation via the large $ N $
limit explicitly leads to  an effective scattering amplitude that
vanishes in the static long-wavelength limit as a consequence of the
crossover to a three dimensional theory for loop momenta $q   \ll  \lambda
T$. This effective scattering amplitude allows us to recognize  that
the effective  {\em three dimensional} 
coupling for soft momenta approaches the three dimensional non-trivial
(Wilson-Fisher) fixed point in the long-wavelength limit near the
critical point. The large $N $ resummation for the relaxation rate
incorporates this effective three dimensional coupling in the spectral
density that determines the imaginary part of the 
retarded self-energy for the order parameter. Since the effective
three dimensional coupling is driven to its fixed point at  
long wavelength,  the contribution from very soft loop momenta $ q \ll
\lambda T$ which give the strongest infrared behavior in lowest order
in perturbation theory is effectively screened by this renormalization
of the coupling. Consequently the most important contribution to the relaxation
rate arises both from the {\em semisoft classical} region of loop
momentum $ T   \gg   q   \gg   \lambda T $ and also from the {\em hard}
region $ q\geq T $. A detailed analysis of the contribution from the
loop momenta reveals a non-perturbative {\bf ultrasoft scale} 
$$ 
k_{us}\simeq{\lambda \; T \over 4 \pi }e^{-{4 \pi\over \lambda}} \; .
$$ 
We find that for soft momenta $ k \gg k_{us} $ the damping rate is
dominated by classical semisoft loop momenta and given by 
\be \label{gamacli}
\Gamma(k,T) \buildrel{k \gg k_{us}}\over= { \lambda T \over 2\pi N }
\left[ 1+ {\cal O}\left( {1 \over \ln{\lambda T \over k}}\right)\right]\; . 
\ee
For $ k \,, m_T \ll k_{us} $ the classical approximation breaks down and the 
damping rate at the critical point $ m_T=0 $  for $ k \ll k_{us} $ is
given by
\be \label{gamaint}
\Gamma(k,T) \buildrel{k \ll k_{us}}\over= { 4\,\pi \, T \over 3\,
N \ln{T\over k}}\left[1+{\cal O}\left({1\over \ln{T\over k}}\right)
\right] \; . 
\ee
For homogeneous  fluctuations near the critical point ($ k=0,~~m_T
\propto |T-T_c|^{1/2}\neq 0$) the damping rate is given by  
\be \label{gamaint2}
 \Gamma_0(m_T,T) \buildrel{m_T \ll k_{us}}\over={ 4\,\pi \, T \over 3\,
N \ln{T\over m_T}}\left[1+{\cal O}\left({1\over \ln{T\over
m_T}}\right) \right]\; .  
\ee
Thus critical slowing down, i.e,  the vanishing of the quasiparticle width $
\Gamma $ for long-wavelengths  emerges in the ultrasoft limit $
k\ll k_{us}  $  or very near the critical point $ m_T \ll k_{us} $ where it
vanishes logarithmically  slow in the $ k, m_T \rightarrow 0 $ limit
to this order in $ 1/N $. Notice that in such regimes the rate is
independent of the coupling $ \lambda $.

The large $ N $ approximation is not limited to 
weak coupling and our results apply just as well to a strong coupling
case $ \lambda \geq 1 $ wherein we find that the  relaxation rate is
given by eq.(\ref{gamaint})-(\ref{gamaint2}). However this analysis
clearly reveals that for weak coupling there 
emerges a {\em hierarchy} of widely separated scales for loop momenta: 
from hard $ q\geq T $ to semisoft $ T   \gg   q   \gg   \lambda T $, and
soft $ \lambda T   \gg   q $ that lead to different contributions to
the relaxation rate. Which is the relevant scale for the damping rate
is determined by the wavevector of the fluctuation of the order
parameter and the proximity to the critical temperature. 
For $ k~,~ m_T \gg k_{us} $ the classical 
approximation does apply
and the damping rate is dominated by the soft and semisoft classical
loop momenta [with the result (\ref{gamacli})], whereas for $ k~,~m_T \ll
k_{us} $ the classical approximation breaks down and the damping rate
is dominated by hard loop momenta $ q\geq T $ [with the results
(\ref{gamaint})-(\ref{gamaint2})].   
  
A similar hierarchy exists in non-abelian
plasmas\cite{lebellac}-\cite{yaffe} and we compare and contrast
our results in the scalar theory with those 
in the hard thermal loop approximation in abelian and non-abelian
plasmas\cite{lebellac}-\cite{yaffe}. 

This article is organized as
follows: in section II 
we introduce the model, obtain the real-time equation of motion for
the order parameter and describe the strategy  followed to obtain the
relaxation rate. In section III we carry out a perturbative analysis
of the relaxation rate to two loops order, 
 recognize the breakdown of perturbation theory and compare to the
case of the hard thermal loop resummation program in gauge
theories. In sections IV and V we introduce the large $ N $ limit, obtain the
effective static scattering amplitude in leading order in the large $
N $ and discuss the dimensional crossover for soft momenta and the
effective three dimensional coupling being driven 
to the three dimensional fixed point. We then use these results to
obtain the relaxation rate and  near criticality to order ${\cal
O}(1/N)$ in the large $ N $ limit and explicitly discuss the screening
of the soft loop momenta. The contribution from classical soft and semisoft
momenta and that of hard loop momenta are analyzed separately to
highlight the important differences. In this section we discuss
further the validity of a quasiparticle interpretation of the
collective long-wavelength fluctuations of the order parameter. In
section VI we summarize our conclusions and results and discuss the
next step of the program. In appendix A the equations of motion in the
large $ N $ limit are derived and in appendix B the polarization integral is
computed. 

\section{Preliminaries: the model and the strategy}

We study the model of scalar fields $ \vec{\Phi}(x) $ in the vector
representation of $ O(N) $, which is conjectured to describe the 
{\em equilibrium} universality class for the chiral phase transition
with two light quarks for $ N=4 $\cite{o4raja}. The Lagrangian density
is given by 
\be
{\cal L}= \frac{1}{2} (\partial_{\mu} \vec{\Phi})^2 -\frac{1}{2}\left[m^2_T+
\delta m^2(T)\right]\; \vec \Phi^2(x)  - \frac{\lambda}{2N} 
[\vec \Phi^2(x) ]^2 +\vec J\cdot \vec \Phi\label{lagrangian}
\ee
\noindent where the external current $\vec J $ has been introduced to
generate an  expectation value for the scalar field (i.e. the order
parameter) by choosing it  to be nonzero along a particular (sigma) direction. 

The counterterm $\delta m^2(T)$ is introduced to cancel the tadpole
contributions (local terms) so that perturbation theory (or the large
$N$ expansion) is carried out in terms of the  effective thermal mass $m_T$. 
In particular to leading order in the large $N$ expansion 
there is the hard thermal loop contribution given 
by the usual tadpole term\cite{parwani} $ \propto \lambda \langle \vec
\Phi^2\rangle/N \propto \lambda \; T^2 $ which combined with the zero
temperature (negative) mass squared leads to an effective finite
temperature  mass $m^2_T\propto (T^2-T^2_c) $. The critical theory
corresponds to  $T=T_c$, i.e. $m_T=0$. In this case the counterterm 
$\delta m^2(T)$ is adjusted consistently order by order to set the
effective finite temperature mass equal to zero. 

As stated in the introduction, our goal is to obtain the relaxation
rate (damping rate) of the order parameter at and near the critical
point. This will be achieved by obtaining the equation of motion for
the expectation value of the scalar field, i.e, the order parameter
and treating its  evolution in real time as an initial value
problem. This is achieved by coupling an external source that serves
the purpose of preparing the initial state. From the equation of 
motion we recognize the self-energy and compute the relaxation rate
from its imaginary part on shell. We write 
\be 
\Phi^a(\vec x,t) = \varphi(\vec x,t)\; \delta^{a,1}+ \eta^a(\vec x,t)
~; ~~ \langle \vec \eta (\vec x,t) \rangle = 0 \label{fieldsplit} 
\ee
\noindent where we chose the particular  direction ``1'' by choosing
external source term to be different from zero along this direction to
give the 
field an expectation value (see below).  The equation of motion for
$ \varphi(\vec x,t) $ is obtained by imposing that  
$ \langle \eta^i(\vec x,t)\rangle = 0 $ consistently in the perturbative
expansion\cite{noneq}. In terms of the spatial Fourier transform of 
the order parameter $ \varphi $ and following the steps detailed in
appendix A.2 (see also\cite{noneq}) we find 
$$
\ddot{\varphi}_k(t)+[k^2+m^2_T +\delta m^2(T)+m^2_{tad}(T)] \; \varphi_k(t) +
\int_{-\infty}^{\infty} \Sigma_{ret,k}(t-t')\;\varphi_k(t')\; dt' = J_k(t)
$$
\noindent where $ J_k(t) $ is the external source that generates the
initial value problem and $ \Sigma_{ret,k}(t-t') $ is
the two-loops retarded self-energy without the tadpole
contributions. The one and two-loops tadpole contributions (local) 
 are accounted for in $ m^2_{tad}(T) $. As described above, the
counterterm $ \delta m^2(T) $ is fixed consistently in perturbation  
theory by requesting that it cancels all constant (in space and time)
contributions to the self-energy (such as the tadpoles) i.e,
$$
 \delta m^2(T_c)+m^2_{tad}(T_c) = 0
$$
The retarded self-energy has a dispersive representation in terms of
the spectral density $ \tilde{\rho}(\omega,k) $ given by 
\be
\Sigma_{ret,k}(t-t')= \int \frac{d\omega'}{2\pi}\;
e^{-i\omega'(t-t')} \int d\omega\;
\frac{\tilde{\rho}(\omega,k)}{\omega-\omega'-i\epsilon}
\label{sigmaretdisprel}  
\ee
\noindent in terms of which the relaxation (damping) rate is given
by\cite{lebellac} 
\be
\Gamma( k,T) = -\frac{\pi}{2\omega_p( k)}\;
\tilde{\rho}(\omega_p(k),k ,T)  \label{dampingrate} 
\ee 
\noindent where $ \omega_p(k) $ is the position of the pole in
the propagator, i.e, the true dispersion relation. For the perturbative two loops  or to leading order in the
large $N$ limit as studied here $ \omega^2_p(k)= k^2+m^2_T $, which at $ T = T_c $
takes the form $ \omega_p(k) = |\vec k | $.

With the purpose of clearly revealing the breakdown of the
perturbative expansion for soft momenta $ k   \ll \lambda T $ we will
begin our analysis by focusing first on the perturbative evaluation of
the damping rate. At one loop order the only contribution to the self
energy is given by the tadpole term $ \lambda \langle \vec \Phi^2(\vec
x,t) \rangle/N $ 
which is local, determines to lowest order  the temperature dependent
mass $m_T \propto |T-T_c|^{1/2}$  and determines the counterterm\cite{parwani}.
 Furthermore this is the leading contribution in the
hard thermal loop limit\cite{lebellac,parwani}.  The lowest order
contribution to the absorptive (imaginary) part of the self-energy
arises at two loops and is studied in detail in the next section. 

\section{Perturbation Theory: two loops}

We begin our study by carrying out a perturbative evaluation of the
damping rate to lowest order, i.e. to two loops to  reveal several
important features of the soft momentum limit, and to pave the way to
implement a non-perturbative evaluation of the self-energy in the
large $ N $ limit. Furthermore, as it will become clear during the
course of the calculation, the lowest order contribution contains some
of the important ingredients of the large $ N $ limit and will
highlight the contribution to the relaxation rate from different
regions of loop momentum.  

After substracting the one and two loops tadpole contributions  which
are cancelled by the counterterm,  the spatial Fourier transform 
of the retarded self-energy reads:
\bea 
\Sigma_{ret,k}(t-t')&=&   8\lambda^2 \; \frac{ N+2}{N^2} \int \dbarp
\dbarq  \left\{G^>_{\vec k +\vec q}(t-t')  \;
 G^>_{\vec p +\vec q}(t-t') \; G^>_{\vec p}(t-t') \right.\nonumber
\\ &-& \left. G^<_{\vec k +\vec q}(t-t')  \;
 G^<_{\vec p +\vec q}(t-t') \; G^<_{\vec p}(t-t')\right\} \Theta(t-t')
\label{twoloopselfenergy} 
\eea
\noindent where the Wightmann functions $G^>, G^<$ are given in Appendix A.1. 

With the purpose of comparing with the results of later sections, it
proves convenient to introduce the intermediate quantities
\bea
{\cal G}^>_{ q}(t-t') & = &  -2\lambda \; \frac{N+2}{N}  \int
\dbarp  \; G^>_{\vec p +\vec q}(t-t') \; G^>_{\vec p}(t-t')\nonumber \\ 
& = & \int dq_0 \; e^{-iq_0(t-t')}\; S^>(q_0,q) \nonumber \\ 
{\cal G}^<_{q}(t-t') & = &  -\frac{2\lambda}{N}(N+2)  \int \dbarp\;
G^<_{\vec p +\vec q}(t-t') \;G^<_{\vec p}(t-t')\nonumber \\ 
& = & \int dq_0 \; e^{-iq_0(t-t')} \; S^<(q_0,q) \nonumber 
\eea
\noindent and using the expression for the Wightmann functions
$ G^>_{\vec k}(t-t'), G^<_{\vec k}(t-t') $ given in  appendix A.1, it is
a straightforward exercise to show that the spectral functions
$ S^<(q_0,q);\; S^>(q_0,q)$ obey the KMS condition
$$
S^<(q_0,q)= e^{-\beta q_0} S^>(q_0,q) 
$$
Introducing the spectral density 
$$
\sigma(q_0,q) = S^>(q_0,q) - S^<(q_0,q)  
$$
which  at the critical point $T=T_c$, i.e. $m_T=0$ is found to be given by 
\bea 
\sigma(q_0,q)= 
 2\lambda\; {N+2\over N} \int \dbarp 
\frac{1}{4\;p |\vec p+ \vec q|} && \left\{ \left[1+n_{\vec q +
\vec p}+n_{\vec p}\right]\left[ \delta(q_0 -|\vec p+ \vec q|-p ) -\delta(q_0
+|\vec p+ \vec q| +p )\right]\right. \nonumber \\
&&\left. +\left[n_{\vec p}-n_{\vec q + \vec p} \right]
\left[ \delta(q_0 -|\vec p+ \vec q|+p ) -\delta(q_0 +|\vec p+ \vec
q|-p )\right] \right\} \label{specdens2lups} 
\eea 
we find
\be
S^>(q_0,q)= \left[1+n(q_0) \right] \; \sigma(q_0,q) \quad, \quad
S^<(q_0,q)= n(q_0) \; \sigma(q_0,q) 
\ee
where
\be
n(q_0)= \frac{1}{e^{\beta q_0}-1}  \quad, \quad 
n_{\vec k} =  {1 \over e^{\beta|\vec k|}-1 }\; . 
\label{boseei}
\ee
The near critical case is considered in sec. III.C.

A lengthy but straighforward calculation with the Bose-Einstein
distribution functions for massless particles leads to 
the following expression for $ \sigma(q_0,q) $
\be
\sigma(q_0,q)= \frac{\lambda}{8\pi^2}{N+2\over N} \left\{
\Theta(|q_0|-q) \; \mbox{sign}(q_0)+ \frac{2T}{q} 
\ln\left[\frac{1-e^{-\frac{|q_0+q|}{2T}}}{1-e^{-\frac{|q_0-q|}{2T}}}\right]
\right\}\;. \label{specdensfin} 
\ee
It is important to emphasize that the second, finite temperature term
is the combination of two different contributions: a two massless
particle cut with support in the region $q_0 > q$ and a Landau damping
cut with support in the region $-q\leq q_0 \leq q$.  

Introducing the Fourier representation of the theta function
\be
\Theta(t-t')= -\int \frac{d\omega}{2\pi
i}\frac{e^{-i\omega(t-t')}}{\omega+i\epsilon} \label{thetafunc} 
\ee 
\noindent in the expression for the self-energy
(\ref{twoloopselfenergy}), we  find the spectral density that enters in 
the dispersive representation of the retarded self-energy
(\ref{sigmaretdisprel}) to be given by 
\bea
\tilde{\rho}(\omega, k) &=& -\frac{4 \lambda}{N} \int \dbarq
\frac{dq_0}{2 |\vec k + \vec q|}\;\sigma(q_0,q)\left[ 
1+n_{\vec q + \vec k}+n(q_0) \right]\left[
\delta(\omega-q_0-|\vec k + \vec q|)-\delta(\omega+q_0+|\vec k + \vec
q|)\right] \nonumber \\
&=& -\frac{4 \lambda}{N} \int \dbarq \frac{1}{2 |\vec k + \vec q|}
\left\{ \sigma(q,\omega-|\vec k + \vec q|)\left[ 
n_{\vec q + \vec k}-n(|\vec k + \vec q|-\omega) \right] \right.\nonumber \\
&+& \left. \sigma(q,\omega+|\vec k + \vec q|)\left[ 
n_{\vec q + \vec k}-n(\omega+|\vec k + \vec q|) \right] \right\}
\label{tilderho}
\eea 
\noindent with $\sigma(q_0,q)$ given by (\ref{specdensfin}) and
we used that,
$$
\sigma(-q_0,q) = -\sigma(q_0,q) \quad, \quad  1+n(-q_0)=-n(q_0)\; . 
$$
Although this seems to be a cumbersome manner
to write down the two loop contribution, it does prove convenient to
establish contact with the large $ N $ description in the next section. 

\bigskip

We are interested in the relaxation of long-wavelength fluctuations of
the order parameter, hence we will consider the soft limit $k \ll T$. 

We study  the contributions
coming from soft  $ q \ll  T $ and hard $q \geq T$ loop-momenta separately. 

\subsection{The soft momenta contribution (classical region)} 

This is the classical region where the Bose-Einstein distribution
functions can be approximated as $ n(\omega) \approx T/\omega, \;
n_{\vec k}\approx T/k  $.

In this regime the contribution of the soft momenta $ q \ll  T $ yields 
$$
\tilde{\rho}_{cl}(\omega, k) = - \frac{\lambda^2 T^2
\omega}{2\pi^2} {N+2\over N^2} \int \dbarq {1 \over q \; |\vec k +
\vec q|^2} \left[ 
{\ln\left|{\omega-|\vec k + \vec q|+q \over \omega-|\vec k + \vec q|
-q}\right|  \over \omega-|\vec k + \vec q|} +
{\ln\left|{\omega+|\vec k + \vec q|+q \over \omega+|\vec k + \vec q|
-q}\right|  \over \omega+|\vec k + \vec q|} \right]
$$
\noindent where we only kept the contribution of order $ T $ to
$\sigma(q_0,q)$.
We evaluate the spectral density at $\omega=k$  leading to the
following form for the soft momenta contribution to the damping rate
(\ref{dampingrate})   
\be
\Gamma_{cl}( k,T) = \left(\frac{\lambda T}{4}\right)^2
\frac{N+2}{2\pi^4 N^2} \left[J_1(k,T)+J_2(k,T)\right]  \label{gamma2} 
\ee
\noindent with $ J_1(k,T) $ and $ J_2(k,T) $ given by the following expressions
\bea
J_1(k,T)&=& \int d^3q ~ \frac{\ln\left|
\frac{|\vec k + \vec q| + q-k}{|\vec k + \vec q| - q-k}
\right|}{q~|\vec k + \vec q|^2~ (|\vec k + \vec q|-k)}
\label{J1} \cr \cr
J_2(k,T)& =&\int d^3q ~ \frac{\ln\left|
\frac{|\vec k + \vec q| + q+k}{|\vec k + \vec q| - q+k}
\right|}{q~|\vec k + \vec q|^2~ (|\vec k + \vec
q|+k)}\label{J2} \; . 
\eea
The angular integrals  can be performed analytically and $k$ can be
scaled out of the integral by introducing the 
variable $ x=q/k $ leading to the final expression 
\be
\Gamma_{cl}(k,T) = \frac{\lambda^2 \; T^2}{16\pi^3 k}{N+2\over
N^2} \int^{\infty}_0 dx \; F[x] \label{gammaint} 
\ee
\indent with  the function $F[x]$  given by 
$$
F[x]= {2\over x}\left[2 \, x \; { \ln x  \over x^2-1}+
\ln\left|{{x+1}\over{x-1}}\right|\right] \; .
$$
\noindent and is depicted in Fig. 1. In principle, the upper limit in the integral in
(\ref{gammaint}) should be $ \alpha T/k $ with $ \alpha \ll 1 $  
to restrict the integral to the soft momenta where the classical
approximation is valid, but the integrand falls off as
$ 1/x^2 $ for $ x   \gg   1 $ and the  
integral is dominated by the small $x$ region $ 0<x \leq 1 $ as shown in
fig. 1. The integral from $ x=0 $ to $ x = \infty $ in
eq.(\ref{gammaint}) gives $ 2\pi^2 $ and the 
contribution from the classical loop momenta $ q   \ll  T $ to the
damping rate to two loops order is thus given by 
\be
\Gamma_{cl}(k,T) = \frac{\lambda^2 \; T^2}{8\pi k}{N+2 \over
N^2}\label{twolooprate} \; . 
\ee
As it will be seen in detail in the next section, the large $ N $
resummation leads to a screening of the soft loop momentum which
cuts off the contribution of momentum $ q< \lambda T $. Hence, with the
purpose of comparing the perturbative two-loop result with that in the
large $ N $ limit, it proves convenient to 
obtain the contribution to the damping rate from the region of semisoft loop
momenta with $ \lambda T \leq q \leq \alpha T $ with 
$ \lambda \ll \alpha \ll 1 $ for the case of soft external
momentum $ \lambda T/k   \gg   1$.  

The contribution to the integral of the function $F[x]$ from this
region is given by  
$$
\int_{\lambda T/k}^{\alpha T/k} dx \; F[x]\approx \frac{4k}{\lambda T}
\ln\frac{\lambda T}{k} 
$$
\noindent therefore this region of loop momentum gives a contribution
to the damping rate given by  
\be
\Gamma_{T   \gg   q > \lambda T} \approx  \frac{\lambda T}{2\pi
}{N+2\over N^2}\ln\frac{\lambda T}{k} \label{largeqdamp} \; . 
\ee
Thus we see that the contribution from the soft region of internal
loop momentum $ q < \lambda T $ contributes a factor 
$ \lambda T/k   \gg   1 $ {\bf larger} than the region of momentum $q>
\lambda T $ for soft external momentum $ k   \ll  \lambda T $. This 
observation will become important when we compare with the large $ N $
result, because as we will show explicitly below, 
the resummation of the effective scattering amplitude will lead to a
softening of the effective vertex and hence screening of very soft
momenta in the loop.  

\subsection{The hard momenta contribution} 

We now focus on obtaining the contribution to the damping rate from
hard loop momenta $ q \geq T $. From the expression of the finite
temperature contribution to  
the spectral density $ \sigma(q_0,q)$ (\ref{specdensfin}) it is
clear that hard momenta $ q_0,q \geq T $ will be exponentially
suppressed {\em unless} either $ |q-q_0|   \ll   T$ or $|q+q_0|   \ll
T $. Consider the expression for the spectral 
density (\ref{tilderho}) for $ \omega =k $ and consider the contribution
from the delta function with support for 
$ q_0= k-|\vec k+\vec q | $; it is straightforward to see that the
other delta function will give a similar contribution.  
For $ q\geq T   \gg   k $ we find that $ |q_0+q| = k|1-\cos \theta | $
where $\theta$ is the angle between $ \vec k $ and $\vec q $ and 
$|q_0-q| \geq 2T$, hence the region of loop momentum that dominates
corresponds to the emission (or absorption) of a pair 
of scalars (the particles in the loop) with total center of mass
momentum {\em collinear} with the external momentum.  

Keeping the leading term $ \propto \ln(|q_0+q|/2T) $ in $ \sigma(q_0,q)
$ and  the {\em full} occupation factors in the expression for 
$ \tilde{\rho} $ the spectral density (\ref{tilderho}) becomes 
$$
\tilde{\rho}(\omega=k, k) \propto -\lambda^2\; T\; k~ \ln{T\over k} 
$$
\noindent and the contribution to the damping rate from the hard loop
momentum region is 
$$
\Gamma_{hard}(k,T) \propto \lambda^2\; T\; \ln{T\over k}
$$
\noindent which is a factor $ (k/T) \ln{T\over k}    \ll
1$ smaller than the  contribution from the 
classical loop momenta (\ref{twolooprate}) for  $ k/T   \ll  1 $.  In
summary, when the external momentum is soft $ k   \ll  \lambda T $
the damping rate is completely determined by the classical region $ q
\ll  T $ of loop momenta and given by eq.(\ref{twolooprate}). That in
a scalar theory the leading temperature effects are determined by the
classical region of loop momenta was already anticipated in
references\cite{aarts,jakovac} but the computation above identifies
the contribution to the relaxation rate from several different regions of loop
momentum. This identification will become important to understand the
result obtained from the large $ N $ limit.  

\subsection{Near criticality}

A calculation very similar to that in the critical inhomogeneous case
can be carried out for homogeneous fluctuations ($ k=0 $)  near
 the critical point with the effective thermal mass $ m^2_T \propto
\lambda(T^2- T^2_c) $ straightforwardly. In this case the angular
integrals are trivial and most of the steps are similar to the
critical case leading to (see also\cite{aarts,jakovac}) 
$$
\Gamma_0(m_T,T) = \frac{\lambda^2 \; T^2}{8\pi m_T}{N+2 \over N^2}
$$
A similar analysis for contributions from different regions of loop
momentum is obtained by replacing $k \rightarrow m_T$ in the arguments above. 

The resonance parameter $ \Gamma(k,T)/\omega_p $ with $ \omega_p $ the
position of the single (quasi) particle 
pole determines how broad is the resonance. If $\Gamma(k,T)   \ll  \omega_p$
the quasiparticle can be described by a narrow resonance and its decay
occurs on time scales much longer than those of the microscopic 
oscillations $\omega^{-1}_p$. On the other hand for $ \Gamma(k,T)   \gg
\omega_p$ the notion of quasiparticle is not appropriate and the
excitation is described by a very short lived broad resonance. 

The two loops calculation reveals that at the critical point $m_T=0$
$$
\frac{\Gamma(k,T)}{k} \buildrel{T \gg k}\over={N+2 \over 8\, \pi N^2}
\left(\frac{\lambda T}{k} \right)^2 ~~;
$$
analogously, near criticality for homogeneous fluctuations $ k=0 $  
$$
\frac{\Gamma(m_T,T)}{m_T} \buildrel{T \gg m_T}\over={N+2 \over 8\, \pi N^2}
\left(\frac{\lambda T}{m_T} \right)^2 \; .
$$
\noindent Hence up to this order a quasiparticle interpretation {\bf is not}
reliable for $k~,~ m_T \ll \lambda T$.

Moreover,  for very soft external momenta or very near the critical
temperature, $ k \, ; m_T \ll\lambda T $ the 
perturbative expansion clearly breaks down and a non-perturbative
scheme must be invoked to study the damping rate. 

This situation is similar to that in the hard thermal loops (HTL) program in
the sense that for external momenta $k\ll \lambda T$ (in gauge 
theories $\lambda$ must be replaced by the gauge coupling squared) a
non-perturbative resummation is needed\cite{lebellac,htl}. However, 
here the similarity ends and the major difference with the HTL program
is revealed: whereas in the HTL case the non-perturbative region is
dominated by {\bf hard} internal loop momenta $q\geq T$, the
relaxational dynamics at the critical point is dominated by {\em soft
classical} (and as it will become clear below, also semi-soft) internal loop
momenta $q \ll T$. The difference can also be clearly seen formally by
restoring the $\hbar$ in the 
contributions: the temperature always appears in the combination
$T/\hbar$ (from the distribution functions), in the HTL program the
gauge coupling constant squared $e^2 \rightarrow e^2 \hbar$ (since
this is the loop counting parameter) hence the HTL scale $e^2 T^2
\rightarrow e^2 T^2/\hbar$.  However in the scalar case the loop
counting parameter is $\lambda \rightarrow \lambda \hbar$, hence the
contribution $ \lambda^2 T^2 $ is {\em classical} i.e. independent of
$\hbar$. Therefore whereas in the HTL program perturbation theory
breaks down at a semiclassical scale $k \propto eT/\sqrt{\hbar}$, at the
critical point of a scalar field theory the perturbative expansion 
breaks down at a {\em classical} scale $k \propto \lambda T$. In the
HTL program the damping rate of collective excitations is typically 
of order $e^2 T$ and the quasiparticle poles (plasmons and plasminos)
are of order $\omega_p \propto e T/\sqrt{\hbar}$ hence for 
weak coupling the long-wavelength quasiparticles are always relatively
narrow resonances. This is in striking contrast with the case 
of a critical scalar theory where the long-wavelength excitation of
the order parameter is gapless.

\section{Large N} 

Having recognized the non-perturbative nature of the relaxation for
the long-wavelength components of the order parameter, we seek to 
use a consistent non-perturbative description and  study the
relaxation of the order parameter in the large $ N $ limit. This limit is best studied by
introducing an auxiliary field that replaces the quartic interaction
via a Gaussian integration\cite{losalamos} (Hubbard-Stratonovich
transformation), hence the Lagrangian density becomes
\be
{\cal L} =  \frac{1}{2} (\partial_{\mu} \vec{\Phi})^2  -
\left[\sqrt{\frac{\lambda}{N}}~\chi(x)+{1\over 2}\left(m^2_T+\delta m^2(T)\right)\right] 
\vec \Phi^2(x)  + \frac12 \chi^2(x) +\vec J(x)\cdot \vec
\Phi(x)\; . \label{lagrauxifield} 
\ee 

Before we engage in a study of the damping rate, it is important to
highlight that the large $ N $ expansion effectively 
provides a reorganization of the perturbative series which for example
at leading order and at  zero temperature is akin to the resummation
of the leading logarithms via the renormalization group for the
scattering amplitude. We now study in 
detail this resummation at finite temperature which will reveal  the
screening of the scattering amplitude for soft 
momenta which in turn will be responsible for screening the
infrared behavior of the spectral functions and the damping rate. 

In this section we will focus on the critical theory $ m_T=0 $. 
The analysis of the off-critical case is given in section V. 

\subsection{Effective Scattering Amplitude} 

To leading order in the large $ N $ the two particle to two particle
scattering amplitude is dominated by $ s$-channel exchange and is
completely determined by the propagator of the auxiliary field
$\chi$. Fig. 2 shows the Dyson sum for the propagator 
of the auxiliary field in leading order in the large $ N $ limit and
fig. 3 shows the $s$-channel scattering amplitude in leading order,
the $ t $ and $ u$-channel contributions are subleading. The bubble diagram
which is the building block of the 
propagator of the auxiliary field (and therefore the $ s$-channel
scattering amplitude) is simpler to be calculated in the 
Matsubara formulation of finite temperature field theory with an
external frequency $\nu_n=2\pi n T$ and given by 
\be
I_{bub}(\nu_n,q) = 2 \lambda T\sum_{\nu_m} \int
\frac{d^3p}{(2\pi)^3} \; {1\over\nu^2_m+{\vec p\;}^2}  \; 
{1\over (\nu_m+\nu_n)^2+(\vec p+\vec q)^2} \; .  \label{ibubble}
\ee
To illustrate the resummation in a more clear manner, we focus on the
static limit which is obtained by setting the external Matsubara
frequency to zero. The strongest infrared behavior and leading
contribution in the high temperature 
limit arises from the term $m=0$ in the Matsubara sum, the remaining
spatial momentum integral is carried out leading to 
\be
I_{bub}(0,\vec q) = {\lambda T\over 4 q} \label{bubble}
\ee 
\noindent and  the $s$-channel scattering amplitude in the static limit 
is given by (see Fig. 3)
$$
M(0,\vec k + \frac{\vec q}{2};0,-\vec k + \frac{\vec q}{2}) = {1\over
N} {\lambda \over 1 + I_{bub}(0,\vec q)}\; \; . 
$$
We see that the effective temperature and momentum dependent coupling
constant, defined as the coefficient of $1/N$ in the s-channel scattering
amplitude in the static limit, is given by
\be
\lambda_{eff}(q)={\lambda \over 1+ {\lambda T \over 4 q} } \; . 
\label{lambdaeff} 
\ee
This expression reveals several noteworthy features. Firstly, we see
that at high temperature in the critical region the actual expansion
parameter in the sum of bubbles is $ \lambda T/q $ with $ q $ the spatial 
momentum tranferred into the loop. The factor 
$ T $ is a consequence of the dimensional reduction and the factor
$\lambda T$ can be interpreted as the dimensionful 
three dimensional coupling. Since the expansion is in terms of
dimensionless quantities the factor $q$ in the denominator 
is required for dimensional reasons. In fact this can be understood
via a parallel with the calculation at zero temperature 
in $4-\epsilon$ space-time euclidean dimensions with a coupling
$\lambda T^{\epsilon}$ with T now some dimensionful scale, 
the loop integral for the massless theory produces a factor
$q^{-\epsilon}$ and for $\epsilon=1$ i.e. the three dimensional 
theory one finds the result for the finite temperature loop in the
static limit. Secondly, the expression for the effective 
coupling (\ref{lambdaeff}) is a result of the large $ N $ resummation
to leading order and is the same as that obtained from the solution 
to the renormalization group equation for the running coupling using
the one-loop beta function obtained in the $\epsilon$ 
expansion and setting $\epsilon =1$, i.e. the large $ N $ resummation
is akin to the resummation obtained from the renormalization group in
euclidean field theory, in the sense that the leading order in the
large $ N $ leads to a running 
coupling which is the same as that obtained from the one-loop beta
function. Thirdly, since the effective expansion parameter in the sum
of bubbles  is $ \lambda T/4 q $ it is convenient to introduce the {\em
three dimensional} coupling  $ \lambda_3(q)=  \lambda T/4 q $ and its
effective counterpart 
\be
\lambda_{3,eff}(q) = {\lambda_{3}(q)\over 1+\lambda_{3}(q)}=
\frac{\lambda T}{4q +\lambda T}\; \; . \label{lambda3deff} 
\ee
\noindent The main point is that this effective three-dimensional
coupling is driven to the three-dimensional Wilson-Fisher 
fixed point $\lambda^*=1$ in the soft momentum limit $q\rightarrow 0$,
while the effective {\em four dimensional} coupling (\ref{lambdaeff})
is driven to the trivial fixed point in this  limit. Hence, whereas the
three dimensional coupling 
$\lambda_3(q)=  \lambda T/4 q$ diverges in the $q \rightarrow 0$
limit, the large $ N $ (equivalent to the renormalization group)
effective coupling $\lambda_{3,eff}(q)$ is driven to a finite
fixed point in the soft momentum limit. Therefore 
the large $ N $ resummation is effectively screening the infrared
divergences associated with the soft momentum limit much 
in the same manner as the resummation implied by the renormalization
group within the $\epsilon$ expansion. Obviously, in 
exactly three euclidean dimensions one could hardly justify the
validity of an $\epsilon$ expansion, but the large $ N $ limit 
provides a non-perturbative framework that includes a similar
resummation. The main point of this discussion is  
the realization that the resummation implied by the large $ N $ limit
provides an effective coupling constant that is well 
behaved in the infrared limit, thus leading to the conclusion that the
simple point-like scattering vertex must be 
resummed before attempting to compute the damping rate or any other
transport coefficient near the critical region.  

The analysis in this section reveals the role played by the scale
$\lambda T$: internal loop momenta $q\ll\lambda T$ lead to
non-perturbative contributions, in the weak coupling limit $\lambda
\ll1$ these non-perturbative scales are {\em classical},  
on the other hand for $q  \gg   \lambda T$ the effective couplings
(either four or three dimensional)  are small for weak 
coupling $\lambda$ and the effective vertices coincide with the
bare vertices.  The implications of this discussion will be
important to understand the different contributions to the relaxation
rate.  

\subsection{The relaxation rate} 

As discussed above at leading order in the large $ N $ limit the only
contribution to the scalar self-energy is a tadpole $\propto \lambda
\langle \vec{\Phi}^2 \rangle /N \sim {\cal O}(1)$, which results in the
effective thermal mass $m_T \propto |T-T_c|^{1/2}$ and is
cancelled by the mass counterterm. In this section we consider the theory  at the critical
temperature where the renormalized temperature dependent mass exactly vanishes.

At next-to-leading order ${\cal
O}(1/N)$ the self-energy obtains an absorptive part and is 
given by the diagram shown in fig. 4. 

In appendix A.2 we provide the details necessary to obtain the
retarded self-energy in terms of a dispersion relation 
as in eq.(\ref{sigmaretdisprel}), with the spectral density
\bea
\tilde{\rho}(\omega,k)&=& -\frac{4 \lambda}{N} \int \dbarq
\frac{dq_0}{2|\vec k + \vec q|}\; \rho(q_0,q)\left[
1+n_{\vec q + \vec p}+n(q_0) \right]\left[ \delta(\omega-q_0-|\vec k +
\vec q|)-\delta(\omega+q_0+|\vec k + \vec q|)\right]\nonumber \\
&=&  -\frac{2 \lambda}{N} \int \dbarq \frac{1}{|\vec k + \vec q|}
\left\{ \rho(\omega-|\vec k + \vec q|,q)\left[ 
n_{\vec q + \vec k}-n(|\vec k + \vec q|-\omega) \right] \right.\nonumber \\
&+& \left. \rho(\omega+|\vec k + \vec q|,q)\left[ 
n_{\vec q + \vec k}-n(\omega+|\vec k + \vec q|) \right] \right\}\;.
\label{tilderholargeN} 
\eea
We performed here the integral over $ q_0 $  by using
the delta functions thereby setting the combination $ q_0
+|\vec q+\vec k| =\pm \omega$ for the respective delta functions.

In appendix A.1 we show in detail that 
\be
\rho(q_0,q)= {1\over \pi}~
{\Pi_I(q_0,q)\over\left[1+\Pi_R(q_0,q)\right]^2+\Pi^2_I(q_0,q)}\; ,
\label{specdensauxi}
\ee
\noindent where $ \Pi_I(q_0,q) $ is given by the leading order in the
large $ N $ limit of the two loop spectral density (\ref{specdens2lups}), as  
\bea 
\Pi_I(q_0,q)=  2\lambda  \pi \int \dbarp  \frac{1}{4 \, p |\vec p+
\vec q|} && \left\{ \left[1+n_{\vec q + \vec p}+n_{\vec p}\right]\left[ 
\delta(q_0 -|\vec p+ \vec q|-p ) -\delta(q_0 +|\vec p+ \vec q|+p
)\right]\right.  \nonumber \\
&+&\left. \left[n_{\vec p}-n_{\vec q + \vec p} \right]
\left[ \delta(q_0 -|\vec p+ \vec q|+p )
-\delta(q_0 +|\vec p+ \vec q|-p)\right] \right\}\; .\nonumber
\eea 
 \noindent The first term, proportional to the sum of the occupation
 factors, corresponds to the two particle cut while the 
second term proportional to the difference is obviously only present
 in the medium and corresponds to Landau damping\cite{lebellac}. The
 real and imaginary parts of the polarization of the auxiliary field
 $\Pi(q_0,q)$ are related by a dispersion relation, i.e,  
\be
\Pi_R(q_0,q) = {1\over \pi} \int d\omega \; \Pi_I(q,\omega)\;
 {\cal P}{1\over{\omega-q_0}}\; .\label{realpi} 
\ee
Keeping the leading temperature dependence we obtain
\be 
 \Pi_I(q_0,q) = \frac{\lambda T}{4\pi q} 
\ln\left[\frac{1-e^{-\frac{|q_0+q|}{2T}}}{1-e^{-\frac{|q_0-q|}{2T}}}\right]
 +{\cal O}\left( \lambda \, T^0\right)\; . \label{hiTimagpi}  
\ee
In  appendix B we show explicitly that the leading temperature
dependence  for the real  part of the polarization operator of the
auxiliary field is given by  
\be
 \Pi_R(q_0,q) = \frac{\lambda T}{4
 q}\left[\Theta(q-q_0)-\Theta(-q-q_0)  \right] +{\cal
O}\left( \lambda \, \ln T\right)\label{hiTrealpi}     
\ee
which in the static limit reduces to (\ref{bubble}). 

It is clear from eq.(\ref{hiTimagpi}) that just like in the case of
perturbation theory up to two loops, there are 
two important regions to consider: i) the classical region with $q_0,q
\ll T$ and ii) the hard region with $q_0,q \geq T$ but with either
$|q-q_0| \ll T$ or $|q+q_0| \ll T$ the other regions of hard momentum
being exponentially suppressed. We will be primarily
interested in the case of soft external momentum $k \ll \lambda T$
i.e. long-wavelength fluctuations of the order parameter. 

In the high temperature limit the spectral density $ \rho(q_0,q) $
takes the explicit form 
\bea
\rho(k + |\vec k+ \vec q|,q) &=& {\frac{4q}{\lambda T}
 L_+ \over \pi^2\left[{4q \over \lambda T}+1\right]^2 + L_+^2}\cr \cr
\rho(k - |\vec k+ \vec q|,q) &=& {\frac{4q}{\lambda T} L_- 
\over {\left[{4q \pi \over \lambda T}\right]^2 + L_-^2}}\nonumber
\eea
where we used eqs.(\ref{hiTimagpi})-(\ref{hiTrealpi}) analyzing
carefully the support of the theta functions in $ \Pi_R(q_0,q) $ and
defined
$$
L_{\pm} \equiv \log\left|{k \pm |\vec k+ \vec q| +q \over 
k \pm |\vec k+ \vec q|  - q } \right|.
$$
Introducing in the integral (\ref{tilderholargeN}) the dimensionless
variable $ y = q/k $ and $ x = \cos \theta $ (where $ \theta $ is the
angle between the vectors $ \vec k $ and $ \vec q $) and setting  $
\omega = k $ yields for the damping rate (\ref{dampingrate})
\bea
\Gamma(k,T) &=& \frac{k^2}{\pi ~ N ~ T} \int_0^{\infty}y^3  dy
\int_{-1}^{+1} \frac{dx}{w(x,y)} \left\{ \left[ {1 \over {e^{\beta \, k
\, w(x,y)} - 1}} - {1\over {e^{\beta \, k\, [w(x,y)-1]} - 1}}\right]\;
{L_+(x,y) \over { \pi^2\left[{4\, k \, y \over \lambda T}+1\right]^2  + L_+^2(x,y)}}
\right. \nonumber \\
&+& \left. \left[ {1 \over {e^{\beta \, k \, w(x,y)} - 1}} - {1\over
{e^{\beta \, k\, [w(x,y)+1]}} - 1} \right] \;
{L_-(x,y) \over { \left[{4\, k \, y \pi \over \lambda T}\right]^2  +
L_-^2(x,y)}} \right\}\label{gama}
\eea
where
$$
L_{\pm}(x,y) = \log\left|{1 \mp w(x,y) +y \over 1 \mp w(x,y) -
y}\right|  \quad \mbox{and} \quad  w(x,y) \equiv \sqrt{1+y^2 + 2xy}\;.
$$
For $\lambda T/ 4k \gg 1$ the region of small $y$ (small momentum) is
screened by the resummation of the scattering amplitude and leads to a
small contribution to the damping rate of order $k$. For $y > \lambda
T/k$ the screening is not effective and the integrals in eq.(\ref{gama}) are
dominated by the neighborhood of the point $ y = y^* $ at which $
L_{\pm}^2(x,y) $ is of the same order as the other square in the
denominators, i.e, $ [4ky /\lambda T]^2 $. For large $ y $ we can expand $
w(x,y) $ as follows 
\be \label{wxy}
w(x,y) = y + x + {\cal O}\left({1 \over y} \right) = y \left[ 1 +
{\cal O}\left({1 \over y} \right) \right]
\ee
We thus find
$$
y^* \simeq {\lambda \; T \over 4\pi \; k}\ln{\lambda \; T \over 4\pi \; k}
$$
[For $ k \ll T $ the $ 1 $ in the denominator of the first term of
eq.(\ref{gama}) is irrelevant in the determination of $ y^* $].

The classical approximation for the occupation numbers, i.e, loop momenta $\ll T$ can be used at
$ y = y^* $ provided $ k \; y^*\ll T $ and $ \lambda \ll 1 $, i.e, 
for $k \gg k_{us}$ where 
$$
 k_{us} \equiv { \lambda \; T \over 4 \pi }e^{-{4 \pi\over \lambda}}
$$
is the {\em ultrasoft} scale. 

As it will be discussed in detail below, for $ k \leq  k_{us} $ hard momenta dominate the contributions to the width $
\Gamma(k,T) $. 

We analyze in the subsequent section the width $ \Gamma(k,T) $ in the
two  regimes : soft for which $T\gg k \gg k_{us}$ and ultrasoft $k \ll k_{us}$. 

\subsection{Classical contribution}

To obtain the contribution from classical momenta $q\ll T$ we perform
the following approximations: i) approximate 
$\Pi_I(q_0, q)$ by its limit for $q_0,q \ll T$
$$
\Pi_I(q_0,q) = \frac{\lambda T}{4\pi q} 
\ln\left|\frac{q_0+q}{q_0-q}\right| 
$$
\noindent with the real part given by the leading temperature
contribution (\ref{hiTrealpi}) and ii) approximate the 
Bose-Einstein occupation factors by their classical limit, i.e. in
(\ref{tilderholargeN}) we replace 
$$
1+n_{\vec q + \vec p}+n(q_0) = T~ \frac{q_0 +|\vec q+\vec p| }{q_0\,
|\vec q+\vec p|} \; . 
$$
We thus find from eq.(\ref{gama}) for the  classical limit
contribution to the damping rate. 
\bea
\Gamma_{cl}(k,T) &=& {k \over \pi \, N } \int_0^{\alpha T/k} y^3 \; dy
\int_{-1}^{+1} {dx \over w^2(x,y)} \left\{ {1 \over 1 - w(x,y)}\;
{L_+(x,y) \over { {1\over\lambda^2_{3,eff}(yk)}+ L_+^2(x,y)}}\right. \cr \cr
&+& \left. {1 \over 1 + w(x,y)}\;
{L_-(x,y) \over { \left[{1\over\lambda_{3,eff}(ky)}-{1\over\lambda^*}\right]^2+
L_-^2(x,y)}} \right\}\label{gamacla}
\eea
\noindent with $\lambda_{3,eff}(q)$ being the {\em effective} three dimensional
coupling given by (\ref{lambda3deff}) and $
\lambda^* = 1 $ the three dimensional fixed point.  We have introduced an
explicit upper momentum cutoff $ q_{max} = \alpha T $ with $
\alpha\ll1 $ that restricts the integration domain to the region where the
classical approximation is valid.

The expression (\ref{gamacla}) clearly  reveals the role of the
three dimensional  
effective coupling $\lambda_{3,eff}(q)$ given
by (\ref{lambda3deff}) and its non-trivial (three dimensional
Wilson-Fisher) fixed 
point $\lambda^*=1$ reached in the soft limit $q \rightarrow 0$.  The
phenomenon of screening of the infrared behavior 
by the renormalization of the coupling is now explicit, the region of
soft loop momentum $q\ll\lambda T$ is independent 
of the coupling and temperature because the effective 
three dimensional coupling is near its non-trivial 
fixed point. The only scale in the integral in the soft-momentum
region is $k$ and a dimensional analysis reveals that 
the contribution from this region is proportional to $k$. 

On the other hand, when the loop momentum is $q  \gg  \lambda T$ the
renormalization 
of the coupling is ineffective and the effective coupling coincides
with the three dimensional coupling $\lambda T/4q$. For weak
coupling $\lambda \ll 1$ and  loop momenta $q  \gg   \lambda T$ the
effective three dimensional coupling is $\lambda_{3,eff}(q)
\approx \lambda_3(q)=\lambda T/4q \ll 1$ and the 
denominators in eq.(\ref{gamacla}) are dominated by the terms
$1/\lambda_{3,eff}(q)$. {\em If} the logarithms can be neglected 
we clearly see that this contribution is the same as that given by the
integrals $ J_1(k,T) $ and $ J_2(k,T) $ given 
by eqs. (\ref{J1}) in the two loop computation of the
damping rate (\ref{gamma2}), which is proportional to 
$\lambda^2 T^2/k$. This region begins to dominate for $q >
\lambda T \ln(\lambda T/k) $ when  $ 1/\lambda_{3,eff}(q) $ becomes 
larger than  the logarithm in the denominators in both integrals in
eq.(\ref{gamacla}). If the loop momenta are such that $ T\gg q $
 in this region,
i.e. the classical approximation is valid, the contribution 
of this region to the integral can be estimated by cutting off the
integrals in eq.(\ref{gamacla}) at a lower momentum of order 
$ q_{min} \approx \lambda T \ln(\lambda T/k) $. Hence following the same
arguments as for the two loops case that led to 
the estimate (\ref{largeqdamp}) we conclude that the region of {\em
classical semi-soft} loop momentum $ T  \gg  q  \gg  q_{min} $ leads to
a contribution to the damping rate  $ \approx \lambda T $.  
However as $ k $ becomes smaller, $ q_{min} $ approaches the cutoff
$\alpha T$, i.e, the limit of validity of the classical approximation and the
logarithmic terms cannot be neglected. In particular for $k \leq
k_{us}$ with $k_{us}$ the ultrasoft scale introduced above 
 the integral becomes sensitive to momenta of order of $T$ and the
classical approximation breaks down. For these ultrasoft momenta of
the fluctuations the damping rate is determined by the region of hard
loop momentum $ q\geq T $.   

This  analysis yields to a preliminary assessment of how  
different regions of loop momentum will contribute to the damping
rate: 

i) the soft region of loop momentum $q\ll\lambda T$ is dominated by
the three dimensional fixed point and contributes to the damping rate  
$$
\Gamma_{q\ll\lambda T} \propto k 
$$
\noindent 
ii) the semisoft region of loop momentum $T  \gg  q  \gg  \lambda T$
is still dominated by classical modes but the renormalization 
of the scattering amplitude is irrelevant. If  $ q_{min} \approx
\lambda T \ln(\lambda T/k) \ll T$ the logarithms can be neglected  and
the integrals in eq.(\ref{gamacla})  behave similarly to the
perturbative two loops 
computation. For momenta $ T  \gg  q  \gg  q_{min} \sim \lambda T
\ln(\lambda T/k) $ the integrals are dominated by the terms in the
denominators proportional to $ 1/\lambda^2_3(q) $  leading to a
contribution 
$$
\Gamma_{T  \gg  q  \gg  q_{min}} \propto \lambda T  
$$
The validity  of the classical approximation and the dominance of 
semisoft loop momenta is warranted for weak coupling when there is a
clear separation between the hard scales with   
 $q \geq  T$ the
semisoft scales with $T  \gg  q \geq \lambda T$ and the soft scales
for which $ \lambda T   \gg   q$.  

iii) However when $ q_{min} \geq T$ the logarithmic terms are the
dominant terms in the denominators  of the integrands and this region
is sensitive to the hard loop momenta $q\geq T$ and the classical approximation is not
warranted. This region requires the full Bose-Einstein distributions and will  be
studied in detail below. 

After this preliminary assessment, we now provide a quantitative analysis of the different regions. 

As argued above in the region $ y\ll\lambda_3(k)=\lambda T/(4 k), $ i.e. $
q\ll\lambda T $, the integrals in eq.(\ref{gamacla}) are independent
of $ \lambda_3(k) $ (i.e, of both the coupling constant and
temperature)  and infrared finite, contributing to the damping rate a  
term that is proportional to  $k$. This is the region of loop momenta
 for which the effective coupling 
is near the three dimensional Wilson-Fisher fixed point.  The
 screening of the scattering amplitude is ineffective for 
loop momenta $q \geq \lambda T$ i.e, the semisoft scales, in this
 region $ y   \gg   \lambda_3(k)\; 
\ln\lambda_3(k) $ and  the term $ [y\lambda_3(k)]^2 $  dominates over the
logarithms leading  to a contribution to the damping rate  
proportional to $ k \lambda_3(k) = \lambda T /4 $. Since in this
 region $ y> \lambda_3(k)  \gg  1$   
we can  approximate $ w(x,y) $ according to eq.(\ref{wxy}) and the
 integrals simplify considerably. 

Up to $ 1/\lambda_3(k) $ corrections we find
\bea\label{aprocla}
\Gamma_{cl}(k,T) &=& {k \over N \pi}\int^{\alpha T/k}_{\lambda_3(k)}
dy  \int_{-1}^{+1} dx\left\{ { \ln{2y \over 1-x } \over \pi^2\left[{y\over
 \lambda_3(k)}+1\right]^2+\ln^2{2y \over 1+x }} \right.\cr \cr
&+&\left. { \ln{2y \over 1+x } \over {\pi^2 \,  y^2\over
 \lambda_3^2(k)}+\ln^2{2y \over 1+x }}\right\} \left[ 1+ {\cal
O}\left(\frac1{\lambda_3(k)}\right) \right] \; .
\eea 
Now the angular integrals (over the variable $x$) can be
performed changing the integration variables to $ y = \lambda_3(k) u $
and expanding in inverse powers of $ \ln \lambda_3(k) $, since for 
$ \lambda \ll \alpha \ll 1 $
$$
\ln\left[{2\, \lambda_3(k) \, u \over 1 \pm x }\right] = \ln \left[2\,
\lambda_3(k)\right] \left[1+{\cal O}\left( {1 \over \ln
\lambda_3(k)}\right)\right] \; . 
$$ 
This is certainly  a slowly converging approximation 
for soft momenta but numerical calculations show it to be  reliable
(see figs. 5-7 below). Then we find from eq.(\ref{aprocla}) 
\bea
\Gamma_{cl}(k,T)&=& { 2 k \; \lambda_3(k) \over N \, \pi} \int_1^{\infty} du
\left[ { \ln[2\lambda_3(k)] \over\pi^2 (u+1)^2 +
 \ln^2[2\lambda_3(k) ] }+{ \ln[2\lambda_3(k)] \over (\pi \, u)^2 + 
\ln^2[2\lambda_3(k)] } \right] \left[ 1+ {\cal
O}\left( {1 \over \ln \lambda_3(k)}\right) \right] \cr \cr
&=& {\lambda T \over 2N\pi} \left[ 1+ {\cal O}\left( {1 \over \ln
\lambda_3(k)}\right)\right]\; . \label{gamaclf}
\eea  
A detailed analysis of both integrals above reveal the presence of
{\em two important scales}: a) the cutoff scale 
$ y  = \alpha T/k $, that is  $ q= \alpha T $ with $ \alpha \ll 1 $
that determines the regime of validity of the classical approximation when the
Bose-Einstein occupation factors may be replaced by their classical
counterparts and b) a scale $ y^* \simeq \lambda_3(k) \, \ln
\lambda_3(k) + \cdots$ at which there is a crossover of 
behavior in the denominators of the integrals. 
 For $ y<y^*, \; y/\lambda_3(k) \ll \ln y $ and the
denominator is dominated by the logarithm, whereas for $ y\gg y^* $,
$y/\lambda_3(k) \gg \ln y $ and the integrands 
behave as $ \ln y /y^2 $ which
is the same behavior as that in the integrals $J _1(k,T) $ and $
J_2(k,T) $ in eqs.(\ref{J1}) for the two loop computation. In the case when
$y^* \ll \alpha T/k$ the integrand falls off very fast and the
integral is independent of the upper cutoff. 

The result (\ref{gamaclf}) is confirmed by a careful numerical study
of the integrals in this range and displayed in figs. 5-7.  
 
The condition that $ y^* \ll \alpha T/k $ translates into the following
condition for $k$  
$$
k \gg \lambda T e^{-\frac{\alpha}{\lambda}} 
$$
It is clear that  $y^*$ becomes of the order of $T/k$ and therefore
the crossover scale becomes of order $T$ for $ k\sim k_{us} \simeq {
\lambda \; T \over 4 \pi }e^{-{4 \pi\over \lambda}} $. Hence for
wavevectors $ k \gg k_{us} $ the classical approximation is valid and
the damping rate is dominated by semisoft classical loop momenta $
T\gg q \gg \lambda T $ and given by    
\be\label{gammafinclaslargeN}
\Gamma_{cl}(k,T) = \frac{\lambda T}{2N\pi}\left[ 1+ {\cal O}\left( {1
\over \ln{\lambda T \over k}}\right)\right] \quad ;  ~~\mbox{for}~~ k
\gg  { \lambda \; T \over 4 \pi }e^{-{4 \pi\over \lambda}} \; .
\ee
In the opposite limit, i.e. for $ k \ll k_{us} $ the crossover scale
$ y^* \gg \alpha T/k $ 
and the terms $ [ y/\lambda_3(k)]^2 $ in the denominators are negligible
as compared to $ (\ln y)^2 $ in this range.   In this case
the integrals can be evaluated by neglecting the $ [y/\lambda_3(k)]^2 $
in the denominators with the result 
$$
\Gamma_{cl}(k,T)\sim \frac{\alpha T}{\ln(\alpha T/k)}\; . 
$$
This cutoff dependence signals the {\em breakdown of the classical
approximation} since the integrand is sensitive to 
hard momenta of order $q \geq \alpha T$. For $ k \ll
k_{us} $  the crossover scale $y^*$ becomes of order $ T/k $, i.e, $ q
 \sim  T $ and we must keep the full occupation numbers, this is the
regime dominated by the hard loop momentum, which is studied
below.   

Thus we conclude that the non-perturbative region of wavevectors for
which the {\em classical} approximation is valid 
is $\lambda T \gg k \gg { \lambda \; T \over 4 \pi }e^{-  4 \pi\over
\lambda} $ and in this region the relaxation rate is given by 
(\ref{gammafinclaslargeN}).
However for long-wavelength fluctuations with wavevectors $ k \ll
k_{us} \sim  { \lambda \; T \over 4 \pi }e^{-{4 \pi\over
\lambda}} $ the classical approximation breaks down and we must
consider the contribution from the hard loop momenta.  

This analysis of the classical contribution reveals that i) the screening of 
loop momenta $q \ll \lambda T$ by the infrared renormalization of the
scattering amplitude makes the damping rate 
a factor $k/\lambda T \ll 1$ {\em smaller} than the lowest order (two
loops) computation,  ii) the damping rate is independent of momentum for $
k \gg { \lambda \; T \over 4 \pi } e^{-{4\pi/\lambda}} $ and given by
en. (\ref{gammafinclaslargeN}) i.e. 
there is no critical slowing down in the regime of validity of the
classical approximation to this order in the large $N$ expansion.

\subsection{Ultrasoft Scale: $ k \ll { \lambda \; T \over 4 \pi } \,
e^{-{4\pi/\lambda}} $} 

We now focus on the computation of the damping rate in the regime of ultra-soft 
fluctuations of the order parameter, i.e,  $ k \ll { \lambda \; T \over 4 \pi } \, e^{-4\pi/\lambda} $. 

In this limit  we expand the {\em difference} of the occupation numbers to order $ k \over T $ {\bf
inside} the integrand in  eq.(\ref{gama}) 
$$
{1 \over {e^{\beta \, k
\, w(x,y)} - 1}} - {1\over {e^{\beta \, k\, [w(x,y)\mp 1]} - 1}} = \mp 
{ k \over T}{e^{\beta \, k\, w(x,y)}\over {\left( e^{\beta \, k\,
w(x,y)} - 1\right)^2}} + {\cal O}\left(  { k \over T}   \right)^2
$$
Since in this region the logarithm dominates, we neglect the terms $ 4
\pi q / \lambda T $ in the denominators. We thus find
\bea\label{gamaus}
\Gamma(k,T)&\buildrel{T \gg k}\over=& {k^3 \over N \, \pi \, T^2} 
\int_0^{\infty}y^3 \; dy \int_{-1}^{+1} {dx \over w(x,y)} 
{e^{\beta \, k\, w(x,y)}\over {\left( e^{\beta \, k\, w(x,y)} -
1\right)^2}}\;\times \cr \cr
&&\left[  -{L_+(x,y) \over { \pi^2 + L_+^2(x,y)}} + {1  \over L_-(x,y)}
\right] \; .\nonumber
\eea
In order to perform the integration it is convenient to change  variables to
$$
v \equiv { k \over T}[w(x,y)- 1] \quad , \quad \sigma \equiv { 2k \over T}
{ w(x,y)- 1 \over y + 1 - w(x,y)}\; .
$$
The width then takes the form
\bea
\Gamma(k,T)&\buildrel{T \gg k}\over=&{ 2 T \over N \pi} \left[
\int_0^{\infty} {e^v \, v^3 \, dv \over \left( e^v -1 \right)^2}
\int_v^{\infty} {d \sigma \over \sigma^2} { \ln \left(1 + {T
\sigma\over k}\right) \over \pi^2 +\ln^2\left(1 + {T \sigma\over
k}\right)} \right. \cr \cr
&+&\left. \int_{2k/T}^{\infty} {e^v \, v^3 \, dv \over \left( e^v -1 \right)^2}
\int_v^{\infty} {d \sigma \over \sigma^2}{ 1 \over \ln \left({T
\sigma\over k}-1\right)}\right]\left[ 1 + {\cal O}\left(  { k \over T}
\right) \right]\; .\nonumber
\eea
We can further approximate these expressions by expanding in inverse
powers of $ \ln {T \over k}  $.  We set,
\bea
&&\int_v^{\infty} {d \sigma \over \sigma^2} { \ln \left(1 + {T
\sigma\over k}\right) \over \pi^2 +\ln^2\left(1 + {T \sigma\over
k}\right)} = \int_v^{\infty} {d \sigma \over \sigma^2}{
1 \over \ln{T\sigma\over k} }\left[ 1 +  {\cal O}\left({1  \over
\ln{ T  \over k}}\right) \right] = { 1 \over v \ln{ T  \over k}}\left[
1 +  {\cal O}\left({1  \over \ln{ T  \over k}}\right) \right] \cr \cr 
&&\int_v^{\infty} {d \sigma \over \sigma^2}{ 1 \over \ln \left({T
\sigma\over k} -  1\right)} = { 1 \over v \ln{ T  \over k}}\left[ 1 +
{\cal O}\left({1  \over \ln{ T  \over k}}\right) \right]\; .\nonumber
\eea
The asymptotic form of the width thus becomes
\be\label{gamaultra}
\Gamma(k,T)\buildrel{k \leq  k_{us}}\over= { 4\,\pi \, T \over 3\,
N \ln{  T  \over 
k}}\left[1+{\cal O}\left({1\over \ln{T\over k}}\right) \right] \; ,
\ee
where we used the integral\cite{grad}
$$
\int_0^{\infty} {e^v \, v^2 \, dv \over \left( e^v -1 \right)^2} = {
\pi^2 \over 3}\; .
$$

There are two important noteworthy features of this result: 
i) the damping rate for ultrasoft fluctuations is
{\em independent of the coupling} and ii) critical slowing down 
of long-wavelength fluctuations emerges in the
{\em ultrasoft momentum} limit with the damping rate vanishing 
only logarithmically as $k \rightarrow 0$. 

The intermediate regime between the soft and ultrasoft scales 
is difficult to study analytically, we therefore studied the
damping rate in a wide range of momentum $k$ numerically. 

Figs. 5-7 display the dimensionless ratio $ N \, \Gamma(k,T)/T $ as
a function of $ \ln{ T \over k} $ for three fixed values of the
coupling: $ \lambda = 12.0 , \; 1.0 $ and $ 0.2 $ respectively  as obtained via a numerical integration of
eq. (\ref{gama}). We see that the damping rate is a monotonically
decreasing function of $ T \over k $ for large enough  values of $ T
\over k $. The smaller is the coupling, the slower $ N \,
\Gamma(k,T)/T $  decreases as a function of $ T \over k $. Furthermore we have established numerically the
reliability of the results in the soft and ultrasoft regimes, thus confirming our detailed analysis in these
cases.  

\vspace{2mm}

\section{Near critical regime: $ \vec k=0,\; |T- T_c| \ll T_c$}

Having understood in detail the critical case we are now in position
to complete the 
study of relaxation by considering the near-critical case,
i.e. $|T-T_c|\ll T_c$. The  
general case of $\vec k \neq 0, T\neq T_c$ is rather complicated but
we can learn much 
by focusing on the homogeneous case $\vec k=0$. There are two
important modifications of the 
previous results that are required to study in detail the near-critical case: 
 
\vspace{2mm}

{\bf i:) } To leading  order  in the large $N$ limit, 
 the finite temperature effective mass squared is given by 
$$
m^2_T \propto \lambda \; (T^2 -T^2_c) 
$$
\noindent therefore the effective mass (inverse of the correlation
length) vanishes near the critical temperature as $m_T \propto
|T-T_c|^{1/2}$ which is the mean-field behavior consistent. Since the
absorptive part of the self-energy is next to leading order ${\cal
O}(1/N)$ we consistently use this effective mass near the critical
point. Therefore to this order the frequencies are given by $
\omega_{\vec p}= \sqrt{{\vec p\;}^2+m^2_T} $.

\vspace{2mm}

{\bf ii:)} The effective static scattering amplitude can be obtained
by replacing the massless Matsubara propagators 
in (\ref{ibubble}) by the corresponding massive ones, and is now given by

$$
\lambda_{eff}\left(\frac{q}{m_T},\frac{T}{m_T}\right)= 
\frac{\lambda}{1+\frac{\lambda T }{2\pi
q}\mbox{arctg}\left[\frac{q}{2m_T}\right]} 
$$
\noindent which now reveals the vanishing of the effective coupling
\be\label{effcoupmass} 
\lambda_{eff}\left(0,\frac{T}{m_T}\right)=
\frac{\lambda}{1+\frac{\lambda T }{4\pi \, m_T}}
\ee
for $ m_T \rightarrow 0 $ [compare with eq.(\ref{lambdaeff}].  

The critical region of relevance corresponds to $m_T \ll \lambda T$
for $T \rightarrow T_c^+$. In the case  
under consideration, for homogeneous fluctuations of the order
parameter ($\vec k=0$) the only dimensionful  
quantity is the effective mass $m_T$ which regulates the infrared
behavior of the integrals. Thus, just as in the 
critical case studied above two different regimes emerge which we
refer to as: {\bf i)} the semicritical regime $m_T\ll\lambda
T/(4\pi)$; {\bf ii)} the ultracritical regime $m_T\ll\lambda
T/(4\pi)e^{-4\pi/\lambda}$. It will become clear below that the
semicritical and the ultracritical regimes correspond respectively to
the soft and the ultrasoft regimes discussed at $T=T_c$,
$k\neq0$. Since the relevant loop momenta are semisoft, $T\gg q \gg
\lambda T$ and hard $q\geq T$ it is clear that the effective coupling
(\ref{effcoupmass}) behaves just as in the critical case for $\vec k
\neq 0$ studied above since for this range of loop momenta $ q/ m_T
\gg 1 $.  

In order to compute the damping rate we need the general expression
for the resummed spectral density $\tilde\rho(q_0,q)$ in presence
of a non-zero thermal mass, which is now given by [see
eq.(\ref{tilderholargeN})]   
\bea
\tilde{\rho}(\omega, k) &=& -\frac{4 \lambda}{N} \int \dbarq
\frac{dq_0}{2 \omega_{|\vec k + \vec q|}}\;\rho(q_0,q)\left[ 
1+n(\omega_{|\vec q+\vec k|})+n(q_0) \right]\left[
\delta(\omega-q_0-\omega_{|\vec k + \vec q|})-\delta(\omega+q_0+
\omega_{|\vec k + \vec
q|})\right] \nonumber \\
&=& -\frac{4 \lambda}{N} \int \dbarq \frac{1}{2 \omega_{|\vec k + \vec q|}}
\left\{ \rho(\omega-\omega_{|\vec k + \vec q|},q)\left[ 
n(\omega_{|\vec q+\vec k|})-n(\omega_{|\vec k + \vec q|}-\omega) \right] 
\right.\nonumber \\
&+& \left. \rho(\omega+\omega_{|\vec k + \vec q|},q)\left[ 
n(\omega_{|\vec q+\vec k|})-
n(\omega+\omega_{|\vec k + \vec q|}) \right] \right\}\nonumber
\eea 
where $\omega_k^2\equiv m_T^2+k^2$, $n(q_0)$ is the Bose-Einstein
distribution function (\ref{boseei}) 
and $\rho(\omega\pm\omega_{|\vec k + \vec q|},q)$ 
is the massive spectral density at leading order in
the $1/N$ expansion, given by equation (\ref{specdensauxi}) with
the following expression for $\Pi_I(\omega\pm\omega_{|\vec k + \vec q|},q)$ 
\cite{Lawrie}:
\begin{eqnarray}
\Pi_I( \omega,p) & = &
\left\{\frac{\lambda}{8\pi}\sqrt{1-\frac{4m_T^2}
{\omega^2-p^2}}\; \mbox{sgn}(\omega) + \frac{\lambda T}{4\pi p}
\ln\left[\frac{1-e^{-\beta \omega_p^+}} {1-e^{-\beta \omega_p^-}}
\right] \right\} \Theta(\omega^2-p^2-4m_T^2) \nonumber \\
&&+ \frac{\lambda T}{4\pi p} \ln\left[\frac{1-e^{-\beta
\omega_p^+}} {1-e^{-\beta \omega_p^-}}\right] \Theta(p^2-\omega^2)
\label{impar1lup} \\
\omega_p^{\pm}&=&\left|\frac{\omega}{2}\pm \frac{p}{2}
\sqrt{1-\frac{4m_T^2}{\omega^2-p^2}}\right|\, .\nonumber
\end{eqnarray}
We notice that (\ref{impar1lup}) reduces to (\ref{hiTimagpi}) in the
critical limit $ m_T/T\to0 $.  

Keeping the leading correction in $  m_T^2 $ yields in the high
temperature regime,
\bea
\Pi_I( \omega,p) =&& \frac{\lambda T}{4\pi p} \left\{\log\left|{\omega +
p \over \omega - p}\right| -{ 4 \, m_T^2 \; p^2  \,\mbox{sgn}(\omega)
\over ( \omega^2 - p^2 )^2} + \left[ \theta(\omega^2 -p^2 - 4 \, m_T^2 ) -
\theta(\omega^2 -p^2 ) \right]\log\left|{\omega + p \over \omega -
p}\right|\right. \nonumber\\ 
&& \left. + {\cal O}( m_T^4) \right\}\label{cormdom}
\eea
The real part $ \Pi_R( \omega,p) $ can now be obtained via the
dispersion relation (\ref{realpi}),
\bea
\Pi_R(q_0,q) &=& \frac{\lambda T}{4 q}\left\{ \Theta(q-q_0)-\Theta(-q-q_0)
+ { 4 \, m_T^2 \over \pi^2 (q^2 - q_0^2) }\left[ \ln{q^2 \over m_T^2}
+ 1 - { q^2 \over q^2 - q_0^2}\left( { q_0^2 \over q^2 }-1 + \ln{q^2
\over q_0^2} \right) \right] \right\} \nonumber \\
&&+ {\cal O}( m_T^4) \label{cormdomR}
\eea
The damping rate for homogeneous configurations is  given by 
$$
\Gamma_0(m_T,T)=-
\lim_{k\to0}\frac{\pi\tilde\rho(\omega_k,k)}{2\omega_k}=
-\frac{\pi\tilde\rho(m_T,0)}{2m_T} \; ,
$$
\noindent which is now given explicitly by
\bea \label{Gamma0}
\Gamma_0(m_T,T)=\frac{\pi\lambda}{m_T N}\int\frac{d^3q}{(2\pi)^3}&&
\frac1{\omega_q}
\left\{\left[n(\omega_q)-n(\omega_q-m_T)\right]\rho(m_T-\omega_q,q)
\right. \nonumber\\ &&\quad
\left.+\left[n(\omega_q)-n(\omega_q+m_T)\right]\rho(m_T+\omega_q,q)
\right\}\;.
\eea

The integral (\ref{Gamma0}) is much simpler than the analogous
expression for $ k\neq0 $, since the angular integration is
trivial. Nevertheless, a complete evalution of (\ref{Gamma0}) requires
a numerical integration. We refer to figures 8 and 9 for a numerical
evalutation of the dimensionless ratio $ N\Gamma_0(m_T,T)/T $ in the
intermediate  ($\lambda= 1.0 $) and small coupling
regimes ($\lambda=0.2$).  

Just as in the critical case in  the near-critical regime with $ m_T\ll\lambda T/(4\pi) $ 
 the integral (\ref{Gamma0}) is dominated by loop momenta of order $q\geq q^*=
\frac{\lambda T}{4\pi}\ln\frac{\lambda T}{4\pi m_T}\gg\frac{\lambda T}{4\pi}
\gg m_T$ hence we can approximate $ \Pi_I(q_0,q) $ and  $
\Pi_R(q_0,q) $ at $ q_0 = m_T \pm \omega_q \simeq m_T \pm q $ as follows
\bea
\Pi_I(m_T \pm q,q) &=& \pm \frac{\lambda T}{4 \pi q}\left[ \ln{2 q \over
m_T } - 1 + {\cal O}\left( {m_T^2 \over q^2} \right)\right] \cr \cr
\Pi_R(m_T \pm q,q) &=&\frac{\lambda T}{4 q}\left\{
\Theta(q-q_0)-\Theta(-q-q_0) \mp { 2 \; m_T \over \pi^2 \; q}\left[
\ln{q^2 \over m_T^2} + 1\right] + {\cal O}\left( {m_T^2 \over q^2}
\right) \right\} \nonumber
\eea
where we have used eqs.(\ref{cormdom}) and (\ref{cormdomR}). Thus as
 anticipated by the discussion of the effective static scattering
 amplitude in the near critical case (\ref{effcoupmass}) we see that
 the real part of the polarization is indeed similar to the one in the
 critical case for the relevant loop momenta up to corrections of 
order $m_T/q \ll m_T/\lambda T$ in the region of semisoft loop momenta. 

Therefore neglecting terms of order $m_T/\lambda T$ which are negligible 
in the region of interest, we find for the spectral densities
expressions similar to these in the critical case : 
$$   
\rho(m_T + \omega_q,q) =\frac{4q}{\lambda T}\; { L(q) \over 
\pi^2\left[{4q \over \lambda T}+1\right]^2 + L^2(q)}
\quad , \quad 
\rho(m_T-\omega_q,q) = -\frac{4q}{\lambda T}\;{L(q) \over 
{\left[{4q \pi \over \lambda T}\right]^2 +L^2(q)}}\; ,
$$
where we have introduced 
$$
L(q)=\log\frac{2\, q}{e \; m_T}\;.
$$

An analysis similar to that in the critical case reveals that soft 
loop momenta $q \ll \lambda T$ are effectively
screened by the renormalization of the scattering amplitude 
in the near critical region, leading to a contribution of order $m_T$ to the
damping rate. Semisoft and hard loop momenta $q > \lambda T$ are not
screened by the resummation of the scattering 
amplitude and determine the leading contributions to the damping rate. 
 
As argued above, the dominat loop momenta in the integral (\ref{Gamma0}) 
are of order $q\geq q^*= \frac{\lambda T}{4\pi}\ln\frac{\lambda
T}{4\pi m_T}\gg\frac{\lambda T}{4\pi} \gg m_T$ hence we can approximate
$$
n(\omega_q)-n(\omega_q-m_T)\simeq n(\omega_q+m_T)-n(\omega_q)\simeq
n'(q)\; m_T\; ,\quad \omega_q\simeq q
$$
\noindent allowing an  analytical estimate the damping rate from the
approximate expression
\be
\Gamma_0(m_T,T)\buildrel{\lambda T \gg m_T}\over=-
\frac4{NT}\int_a^{\infty} \frac{dq}{2\pi^2}\;q^2 \; n'(q)
\left\{ { L(q) \over 
\pi^2\left[{4q \over \lambda T}+1\right]^2 + L(q)^2}
+{ L(q) \over 
{\left[{4q \pi \over \lambda T}\right]^2 + L(q)^2}}\right\} \label{masainte}\;.
\ee

The resulting integrals are infrared finite but having recognized that the 
leading contribution arises from the
semisoft loop momenta $q\geq \lambda T$ we have 
introduced  an explicit infrared cutoff $ a={\cal C} \; m_T $ with
${\cal C}\gg 1$. Since the integral is dominated by semisoft and hard
loop momenta $ q^*\gg \lambda T $ we can  approximate further $ {4q
\over \lambda T}+1\simeq {4q \over \lambda T} $ whence the two
contributions to the damping rate coincide. Moreover, the dependence
in $a$ is negligible in the critical limit. This is  confirmed  by our
numerical analysis which uses the exact expression (\ref{Gamma0}), the
results of which are displayed in figures 8 and 9. The integrals in
(\ref{masainte}) again reveal a crossover scale $q^*$ at which $4\pi
q^*/\lambda T \sim L(q^*)$. For $q\ll q^*$ the logarithmic term
$L^2(q)$ dominates in the denominators and for $q\gg q^*$ the term
$ (4\pi q^*/\lambda T )^2 $ dominates and the integrand falls off just
as in the perturbative two loops case.  

We now distinguish between the following two possibilities:

\begin{itemize}

\item $\lambda T/(4\pi)e^{-4\pi/\lambda}\ll 
m_T\ll\lambda T/(4\pi)$, to which we refer as the semicritical regime.
In this case $q^*\ll T$ is soft and the classical approximation  to
the Bose-Einstein distribution functions applies. Furthermore, we can
expand in inverse powers of the logarithm
$ \log\lambda_3(m_T)=\log\frac{\lambda T}{4 m_T} $ and a straightforward
analysis along the lines presented for the critical case reveals that
the damping rate is approximatively constant and given by 
$$ 
\Gamma_0(m_T,T)=
\frac{\lambda T}{2\pi N}\left[ 1+ {\cal O}\left( {1 \over \ln
\lambda_3(m_T)}\right)\right]\;.
$$
This is the same result as in the case $T=T_c,\; k\neq0$,
Eq.(\ref{gamaclf}).  This result is of course expected, in the 
semisoft region of loop momentum $T\gg q \gg \lambda T$ and for
$\lambda T/(4\pi)e^{-4\pi/\lambda}\ll m_T\ll\lambda T/(4\pi)$ the
screening of the scattering amplitude is ineffective  and the term
$(q/\lambda T) $ dominates over the logarithms, therefore the
dependence on the mass is negligible in this region.  

\item
$m_T\ll\lambda T/(4\pi)\; e^{-4\pi/\lambda}$, which we refer to as the
ultracritical regime. In this case $q^*\gg T$ is hard and 
the classical approximation breaks down and the full Bose-Einstein
occupation factors must be kept. In this regime  the logarithm gives 
the dominant contribution in the denominators and the asymptotic
damping rate is given by 
\be
\Gamma_0(m_T,T)=
-\frac4{\pi N T}\int_a^\infty dq\; q^2 \; \frac{n'(q)}{\log{2q \over e
\, m_T}}= \frac{4\pi}{3N}\frac T{\log(T/m_T)}
\left[1+{\cal O}\left({1\over \ln{T\over m_T}}\right) \right]\; ,
\label{ultracrit} 
\ee
\noindent which  is exactly the same result as in  equation
(\ref{gamaultra}) with the momentum scale $k$ replaced by the thermal
mass $m_T$. It must be noticed that in this regime there is {\bf no}
dependence on the coupling $\lambda$.

\end{itemize}

\vspace{2mm}

Thus we conclude that the relaxation  rate
near the critical point $ |T- T_c| \ll \lambda T_c $ and $ \vec k = 0 $ has the same
features as the critical rate for $ k \neq 0 $ and $ T = T_c $, provided
we exchange the infrared scales $ m_T $ and $ k $. 

\vspace{2mm}

{\bf Discussion of the results:}

The two loops calculation in  perturbation theory at the critical point 
revealed the importance of the
different scales of loop momentum. The loop integrals are dominated by
the contribution of the soft momentum scales $ q \ll \lambda T $. The
contribution from semisoft loop momenta $T\gg q \gg \lambda T$ is
subdominant by a factor $(k/\lambda T) \ln[\lambda T/k] \ll 1$ in the
long-wavelength  limit $k\ll \lambda T$ and the contribution from hard loop
momentum modes $q \geq T$ is suppressed even further in the weak
coupling limit by an extra power of the coupling $\lambda$.  

The large $ N $ limit leads to a non-perturbative resummation and
results in an infrared renormalization of the static scattering
amplitude as a consequence of {\em dimensional reduction} and crossover to an
effective three dimensional theory for momenta $q \ll \lambda T$. 

The effective three dimensional coupling that emerges from this
analysis of the static scattering amplitude at the critical point is  
$\lambda_{3,eff}(q)=\lambda T/(4q + \lambda T)$ which is driven to the
Wilson-Fisher three dimensional fixed point as $q\rightarrow 0$. Thus
soft loop momenta $q \ll \lambda T$ are effectively screened by
this infrared renormalization of the coupling but semisoft loop
momenta $q > \lambda T$ are coupled with the three
dimensional coupling $\lambda_3(q) = \lambda T/4q$ and  infrared
screening is ineffective for these. The importance of this
effective coupling for the damping rate can be understood intuitively
from figures 2, 3 and 4: the resummation of 
bubbles that leads to the effective scattering amplitude also
renormalizes the spectral density that determines the self-energy, as
shown in figure 4.  

This is   precisely the most important mechanism that leads to our
results in the large $ N $ limit. Whereas the lowest order
perturbative calculation was dominated by the contribution of soft
loop momenta $q \ll \lambda T$ these are effectively screened by the 
infrared renormalization of the coupling which is near the three
dimensional fixed point. The dominant contribution now arises from the
semisoft $q \gg \lambda T$  and hard $q \geq T$ loop momentum. In the
 perturbative computation at two loops these scales
provided subleading contributions of order $ \lambda T $ and $
\lambda^2 T$ respectively to the damping rate. 

Clearly the resummation via the large $ N $ approximation incorporates
the screening of the soft loop momentum scales but also reveals the
emergence of an {\em ultrasoft} scale  $ k_{us} \sim { \lambda \; T \over
4 \pi }e^{-{4 \pi\over \lambda}} $. At the critical temperature for $
k \gg k_{us} $ or  for homogeneous fluctuations  
near the critical point  $T_c \gg m_T \gg k_{us}$ the damping
rate is determined by  
the contribution of the {\em classical, semisoft} loop momentum scales
$ T \gg q \gg \lambda T $, the soft scales  
$ \lambda T \gg q $ being screened by the infrared renormalization of
the coupling and the crossover to a three dimensional 
effective theory. For long-wavelength fluctuations of the order
parameter at the critical temperature with $ k \leq  k_{us} $ or for
homogeneous fluctuations near criticality  
in the {\em ultracritical region}  $m_T \leq k_{us}$  the classical
approximation breaks down and the damping rate is completely
determined by the hard loop momenta $q\geq T$. Critical slowing down, i.e. 
the vanishing of the relaxation rate as $ k\rightarrow 0 $ or for
$k=0$ as  $ T\rightarrow T_c^+$  only emerges in this ultrasoft limit
as shown by equations (\ref{gamaultra}) and (\ref{ultracrit}).   

Thus, whereas the large $ N $ expansion has provided a consistent
resummation and the important ingredient of screening of 
the couplings for the soft loop momentum modes and leads to critical
slowing down of long wavelength fluctuations important 
limitations of  the results obtained here remain. As
we argued in the beginning sections a quasiparticle interpretation of
the long-wavelength collective excitations of the order parameter
requires that the resonance parameter $\Gamma(k,T)/\omega_p(k) \ll 1$
with $\omega_p(k)$ being the position of the quasiparticle pole or 
effectively the microscopic time scale of oscillations of these
fluctuations. To leading order in the large $ N $ limit
$\omega_p(k)=\sqrt{k^2+m^2_T}$ in the 
calculation of the damping rate and the resonance parameter. 
Although critical slowing down emerges from the large $ N $ limit, we
see  that our results  to this order indicate that
$\Gamma(k,T)/\omega_p(k) \gg 1$ for $k, m_T \rightarrow 0$.

There are several possible alternatives: either the quasiparticle
picture is not appropriate to describe the collective fluctuations of 
the order parameter at or near the critical point or further
resummations and or other contributions must be taken into account to
obtain a description of critical slowing down of collective
fluctuations that can be understood within a quasiparticle picture. In
particular an assessment of i) vertex corrections, and ii) wave
function renormalization must be pursued which, however, are beyond
the leading order in the large $ N $ studied here and thus outside the
scope and goals of this article. We are currently studying these contributions
and expect to report our conclusions in a forthcoming article.    

\section{Conclusions and further questions}

In this article we have began the program of studying transport and
relaxation at and near the critical point in second order phase
transitions.  The focus here is to provide a systematic study of
critical slowing down from first principles in a phenomenologically 
motivated quantum field theory. The ultimate  goal of this program  is to 
assess the potential experimental signatures associated with the
critical slowing down of long wavelength fluctuations at the chiral
phase transition. Obtaining a robust understanding of such
phenomena will have important implications in the QGP and or chiral phase
transitions in early universe cosmology and ultrarelativistic heavy
ion collisions.   

Our study reveals novel phenomena that require a non-perturbative
framework for their consistent and sistematic treatment which is
different from the hard thermal loop resummation program used in gauge
theories.   

Whereas critical slowing down has been studied thoroughly in {\em
classical} critical phenomena\cite{MA,hoh} and these 
results were used for  preliminary estimates of the correlation length
at freezeout in heavy ion collisions\cite{rajaslow},
we are  aware of only one prior attempt\cite{pietroni} to study
critical slowing down in a full relativistic quantum field
theory and a similar recent analysis\cite{newrenor}. In
references\cite{pietroni,newrenor}  the Wilson 
renormalization group was used to explore the relaxation of the $ k=0 $
mode of the order parameter slightly away from criticality working at
$ N=1 $. The final results of \cite{pietroni,newrenor} are that for $
t =(T-T_c)/T_c $ approaching the critical limit the damping rate for homogenous
configurations vanishes as $ t^\nu\log t$ with $\nu\simeq 0.5-0.6 $.
The analysis of\cite{pietroni,newrenor}, relies on a truncation of the
exact renormalization group equations and their numerical evolution. 
In our opinion there are two very important limitations in this
approach: i) in refs.\cite{pietroni,newrenor} the absorptive parts were treated
in a rather simplistic manner and associated with the scattering
vertex rather than the self-energy, but more importantly, this
simplified treatment does not include consistently the Landau damping
and multiple particle thresholds that are the important ingredients in
a consistent and sistematic description of damping and relaxation. ii) 
the method relies on the numerical evolution of a set of equations
that had been truncated without a clear control or 
understanding of the errors induced by this truncation. 
In particular, it is not clear if the result $ \Gamma(t) \sim
t^\nu\log t $ is stable with respect to different truncations.

In our study we have systematically focused on the important aspects associated
with Landau damping, many-particle threshold effects and  a consistent
study of real-time phenomena at finite  
temperature. As is evident in our study of absorptive parts of the self-energy
in sections II and III  a simplified treatment that does not include
consistently these can hardly reveal the 
rich hierarchy of scales and the different physics associated with
these: the soft scale  
$ k_s \simeq \lambda T/(4\pi) $ and the ultrasoft scale 
$  k_{us} \simeq\lambda T\exp[-4\pi/(\lambda T)] $. Consistently to
next to leading order in the large $N$ limit  we 
see that  slowing down of relaxation of long-wavelength fluctuations
only begins to emerge at the ultra-soft scale and 
in contrast to the results obtained in\cite{pietroni,newrenor}, we
obtain $ \Gamma \sim T/[N\log t] $. Thus, although 
there is agreement on the statement that the relaxation rate vanishes
at criticality the consistent large $N$ resummation leads to a
very different behavior of the relaxation rate.

\bigskip

Our main results can be summarized as follows: a consistent treatment
of critical slowing down and of transport phenomena at or near a
critical point requires a non-perturbative framework to resum the
contributions from {\em soft} loop momenta which is different from the
hard thermal loop program of abelian and non-abelian gauge plasmas. A
perturbative two loop calculation reveals clearly the emergence of a
hierarchy of loop momentum scales from hard $ q\geq T $ to semisoft $
T \gg q \gg \lambda T$ and soft $ \lambda T \gg q $,  for weak
coupling $ \lambda \ll 1 $ the scales in this hierarchy are widely
separated and the semisoft and soft scales are classical. Recognizing
the shortcomings of  
a perturbative treatment for long-wavelength fluctuations, we
implemented a non-perturbative resummation via the 
large $N$ limit to next to leading order. The large $ N $ limit
provides a consistent non-perturbative framework for 
resummation of infrared contributions. It clearly displays the  
infrared renormalization of the scattering amplitude in the static
limit at or near the critical point and the crossover to three
dimensional physics for soft loop 
momentum. The resummation of the scattering amplitude leads to an
effective three dimensional coupling that interpolates
between the bare coupling for loop momenta $ q \gg \lambda T $ and the
three dimensional Wilson-Fisher fixed point for $ q \ll \lambda T $. 

The infrared renormalization of the effective coupling screens the
contribution from soft loop momentum to the self-energy and the 
relaxation rate, which is now dominated by the contribution of {\em
semisoft and hard loop momenta} $T \gg q \gg \lambda T$. Furthermore a new
{\em ultrasoft} (in the weak coupling limit) non-perturbative
 scale emerges $ k_{us} \simeq { \lambda \; T \over 4 \pi }e^{-{4
\pi\over \lambda}} $ that signals the
breakdown of the classical approximation and the dominance of hard
loop momentum modes.  

For $ k\;, m_T  \gg k_{us} $ the damping rate is dominated by the classical
semisoft scales and given by $ \Gamma(k,T) = \frac{\lambda T}{2N\pi}
$ whereas for $ k \leq  k_{us} $ the hard loop momenta region
dominates and leads to the damping rate $ \Gamma(k,T) = { 4\,\pi \, T
\over 3\, N \ln{T \over k}} $ at criticality or 
$ \Gamma_0(m_T,T) = { 4\,\pi \, T \over 3\, N \ln{T \over m_T}} $  near
criticality for homogeneous fluctuations,  which reveal the slowing
down of relaxation of critical ultrasoft fluctuations with a damping
rate that is {\em independent of the coupling}.    

As discussed above, however, these results and those found
in\cite{pietroni,newrenor} seem to indicate a breadown of the quasiparticle
picture of collective excitations of the order parameter because the
resonance parameter $ \Gamma(k,T)/\omega_p(k) \gg 1 $ in the
long-wavelength limit and the excitation decays on time scales much
shorter than the natural oscillation time $ \omega^{-1}_p(k) $. At this
stage it is not clear if this feature is a true physical manifestation
of relaxation of collective excitations at or near the critical point
or that further resummation  and other contributions that are beyond
the leading order in the large $ N $  must be accounted for. We are
currently studying this possibility by introducing the renormalization
group at finite temperature and analyzing in detail the contribution
from vertex and wave function renormalizations and expect to report on
further understanding on these issues in a forthcoming article. At
this stage our  
study has revealed a wealth of new phenomena and a hierarchy of scales
which will require a deeper understanding for a complete and
consistent treatment of transport and eventually hydrodynamics near or
at the critical point. Only a thorough understanding of these 
phenomena can lead to an unambiguous assessment of the
phenomenological implications of critical fluctuations either in the
formation of cosmological relics in the early universe or in
experimental observables in ultrarelativistic heavy ion collisions
thus motivating and justifying their study.  

\bigskip

{\large\bf Acknowledgements}\\
 
D. B. thanks the N.S.F for partial support through grant awards:
PHY-9605186 and INT-9815064 and 
LPTHE (University of Paris VI and VII) for warm hospitality and
partial support, he also thanks S. Raja for interesting conversations.  H. J. de Vega thanks the Dept. of Physics at the
Univ. of Pittsburgh for hospitality.  We thank the CNRS-NSF
cooperation programme  for partial support. 
The early stages of the work of M.S. were supported by a grant from
Padova University.

\appendix

\section{Equations of motion and spectral densities in the large $ N $ limit}
We study the relaxation of the order parameter via the real time
description of non-equilibrium quantum field theory. This formulation 
requires the time evolved density matrix and is cast in terms of a
path integral along a contour in complex time, a forward branch 
corresponds to the forward time evolution of the density matrix via
the unitary time evolution operator and the backward branch represents
the inverse unitary time evolution that post-multiplies the density
matrix. Consequently there are four propagators: corresponding to 
fields on either branch. For a more complete description of this
formulation the reader is referred to\cite{noneq} and references
therein. The main ingredient in this program are the  free field
Wightmann and Green's functions for the bosonic field $\vec{\Phi}$. In
terms of the spatial Fourier transform of the bosonic fields
$\vec{\Phi}_{\vec k}$ these are given by  
\begin{eqnarray}
&&  \langle \Phi_{\vec k,a}(t) \Phi_{-\vec k,b}(t') \rangle_o \equiv 
\langle \Phi^-_{\vec k,a}(t) \Phi^+_{-\vec k,b}(t') \rangle_o =
-i \; \delta_{a,b} \; G_{\vec k}^{>}(t,t')~, \cr \cr
&&  \langle \Phi_{\vec k,a}(t') \Phi_{-\vec k, b}(t) \rangle_o  \equiv  
\langle \Phi^-_{\vec k,a}(t') \Phi^+_{-\vec k, b}(t) \rangle_o  = -i \; 
\delta_{a,b} \; G_{\vec k}^{<}(t,t')~, \cr \cr
&&\langle \Phi^+_{\vec k,a}(t) \Phi^+_{-\vec k,b}(t') \rangle_o =
-i \;  \delta_{a,b} \; G_{\vec k}^{++}(t,t')~, \cr \cr 
&&\langle \Phi^-_{\vec k,a}(t) \Phi^-_{-\vec k,b}(t') \rangle_o =
-i \;  \delta_{a,b} \; G_{\vec k}^{--}(t,t')~, \label{GMM}\\
&&G_{\vec k}^{++}(t,t')=G_{\vec k}^{>}(t,t') \; \Theta(t-t')+G_{\vec
k}^{<}(t,t')  \; \Theta(t'-t)\, ,\cr \cr 
&&G_{\vec k}^{--}(t,t')=G_{\vec k}^{>}(t,t') \; \Theta(t'-t)+G_{\vec
k}^{<}(t,t') \;  \Theta(t-t')~, \cr \cr 
&&G_{\vec k}^{+-}(t,t')=G_{\vec k}^{<}(t,t') ~,\cr \cr 
&&G_{\vec k}^{-+}(t,t')=G_{\vec k}^{>}(t,t') ~, \cr \cr 
&&G_{\vec k}^{>}(t,t') = G_{\vec k}^{<}(t',t)~. \cr \cr 
\end{eqnarray} 
where $\langle A(t) B(t') \rangle= \mbox{Tr}[ A(t) B(t') \rho(0)]$ denotes the 
expectation value of Heisenberg field operators with respect to
the initial normalized density density matrix which is taken to
describe a thermal state and the subscript $o$ refers to free
fields. It is clear that these real time propagators satisfy the identity:
$$
G_{\vec k}^{++}(t,t')+G_{\vec k}^{--}(t,t')-
G_{\vec k}^{+-}(t,t')-G_{\vec k}^{-+}(t,t')=0~.
$$
The retarded and advanced propagators are defined as
\begin{eqnarray*}
G_{{\rm R},\vec k}(t,t')&=&G_{\vec k}^{++}(t,t')-G_{\vec k}^{+-}(t,t')
=\left[G_{\vec k}^>(t,t')-G_{\vec k}^<(t,t')\right]\Theta(t-t')~,\\
G_{{\rm A},\vec k}(t,t')&=&G_{\vec k}^{++}(t,t')-G_{\vec k}^{-+}(t,t')
=\left[G_{\vec k}^<(t,t')-G_{\vec k}^>(t,t')\right]\Theta(t'-t)~,
\end{eqnarray*}
Where for the cases under consideration with fields in thermal equilibrium 
\begin{eqnarray}
G_{\vec k}^{>}(t,t')&=&\frac{i}{2 k} \left[[1+n_{\vec k}] \; 
e^{-ik(t-t')}+n_{\vec k} \; e^{ik(t- t')} \right]~, \label{ggreat}\\
G_{\vec k}^{<}(t,t')&=&\frac{i}{2 k}\left[
n_{\vec k}~ \; e^{-ik(t- t')} +[1+n_{\vec k}]e^{ik(t-t')} \right]~,
\label{gsmall}\\
n_{\vec k}& = & {1 \over \exp(\beta\omega_{\vec k})-1 }~ . \nonumber
\end{eqnarray}
From the Lagrangian density in terms of the auxiliary fields given by
 (\ref{lagrauxifield}) it is straightforward to find the 
 free field real time correlation functions for the auxiliary
 fields. In terms of the spatial Fourier transform of the 
auxiliary field $\chi_{\vec k}$ these are given by 
\bea
&& \langle \chi^+_{\vec k}(t) \chi^+_{-\vec k}(t') \rangle_o =
 i\delta(t-t') \cr \cr 
&& \langle \chi^-_{\vec k}(t) \chi^-_{-\vec k}(t') \rangle_o =
 -i\delta(t-t') \cr \cr 
&& \langle \chi^+_{\vec k}(t) \chi^-_{-\vec k}(t') \rangle_o =
 \langle \chi^-_{\vec k}(t) \chi^+_{-\vec k}(t') \rangle_o =0 \nonumber
 \eea
Figures 10(a) and 10(b) depict the series of Feynman diagrams for the {\em full} $\langle \chi^+ \chi^+ \rangle$ propagator and for the
{\em full} plus-plus component of the propagator of the composite field $(\vec \Phi)^2 = \vec \Phi \cdot \vec \Phi$, i.e. $\langle (\vec \Phi^+)^2 (\vec \Phi^+)^2 \rangle$. Figures 11(a),  11(b) and 12(a), 12(b) depict similar relations for the $\langle \chi^+ \chi^- \rangle$
and $\langle \chi^- \chi^- \rangle$ propagators. 
Thus using the free field propagators for the auxiliary field given
above we find the following 
relations to {\em all orders} for the {\em full} propagators
\bea
&& \langle \chi^+_{\vec k}(t) \chi^+_{-\vec k}(t')\rangle =
i\delta(t-t') +\lambda \langle \Phi^{+,2}_{\vec k}(t)
\Phi^{+,2}_{-\vec k}(t') \rangle \label{PPrelation} \\ 
&& \langle \chi^+_{\vec k}(t) \chi^-_{-\vec k}(t')\rangle = \lambda
\langle \Phi^{+,2}_{\vec k}(t) \Phi^{-,2}_{-\vec k}(t') \rangle
\label{PMrelation} \\ 
&& \langle \chi^-_{\vec k}(t) \chi^-_{-\vec k}(t')\rangle =
-i\delta(t-t') +\lambda \langle \Phi^{-,2}_{\vec k}(t)
\Phi^{-,2}_{-\vec k}(t') \rangle \label{MMrelation}  
\eea
\noindent with the definition
$$
\Phi^{\pm ,2}_{\vec k}(t) \equiv \int d^3 x~ e^{i\vec k \cdot \vec x}~
\vec \Phi^{\pm}(\vec x,t) \cdot \vec \Phi^{\pm}(\vec x,t) 
$$
The correlation functions of the bilinear composite operator can be
written in terms of spectral densities in the following manner 
\bea
&&\langle \Phi^{+,2}_{\vec k}(t) \; \Phi^{+,2}_{-\vec k}(t') \rangle =
\int d\omega \left[\rho^>_{\phi^2 \phi^2}(\omega; k) \; \theta(t-t')+
\rho^<_{\phi^2 \phi^2}(\omega; k) \; \theta(t'-t)\right]
e^{-i\omega(t-t')} \label{rhogreatphi2} \\ 
&&\langle \Phi^{+,2}_{\vec k}(t) \Phi^{-,2}_{-\vec k}(t') \rangle =
\int d\omega   \;   \rho^<_{\phi^2 \phi^2}(\omega; k)  \; 
 e^{-i\omega(t-t')} \label{rhopmphi2}
\eea 
Familiar manipulations introducing a complete set of
energy eigenstates in the trace lead to the KMS condition
\be
\rho^<_{\phi^2 \phi^2}(\omega; k) = e^{\beta \omega} \; 
\rho^>_{\phi^2 \phi^2}(\omega; k) \label{KMS2} 
\ee
The main reason for presenting these formal steps is that the
auxiliary field itself {\em does not} have a KMS relationship for 
its spectral functions because it is not a canonical field but a
Lagrange multiplier. However the  relations 
(\ref{PPrelation}-\ref{MMrelation}) which hold to {\em all orders} relate the correlators of the
auxiliary field to those of the bilinear composite 
operator for which the spectral functions associated with their
correlators do obey the KMS condition.  

Writing the retarded correlator for the auxiliary fields as a spectral
representation  
$$
\langle \chi^+_{\vec k}(t) \chi^+_{-\vec k}(t')\rangle - \langle
\chi^+_{\vec k}(t) \chi^-_{-\vec k}(t')\rangle = 
i \int \frac{dq_0}{2\pi} \;  \rho_{\chi}(q_0,k) \;  e^{-iq_0(t-t')},
$$
\noindent using the spectral representations
(\ref{rhogreatphi2})-(\ref{rhopmphi2}) and  the representation for
$ \Theta(t-t') $ given by (\ref{thetafunc}) we obtain the relation
between the spectral representation for the retarded correlator of the
auxiliary field and that for the bilinear composite in the following form
\be
\rho_{\chi}(q_0,k)= 1 + \lambda \int d\omega~ \rho^>_{\phi^2
\phi^2}(\omega;k)~   \; 
\frac{1-e^{-\beta \omega}}{q_0-\omega +i\epsilon}
\label{rhosrelation} 
\ee
\noindent where we have used the KMS condition (\ref{KMS2}). The next
step of the program is to obtain the spectral density
$ \rho_{\chi}(q_0,k) $ to leading order in the large $ N $
limit. This is achieved through linear response analysis for the 
expectation value of the auxiliary field.

\subsection{Linear response for the auxiliary field} 

The real time expectation value of the 
auxiliary field is obtained by coupling an external source to the 
auxiliary field in the original
Lagrangian (\ref{lagrauxifield}) ${\cal L} \rightarrow 
{\cal L}+J_{\chi}\chi$ with the {\em same} external source for the two
time branches. Assuming the addition of counterterms in the
Lagrangian to ensure that the expectation value of the auxiliary field 
vanishes for vanishing external source we have
$$
\delta_{\vec k}(t) \equiv \langle \chi^+_{\vec k}(t) \rangle = i \int
dt' J_{\chi ,\vec k}(t') \left[ \langle \chi^+_{\vec k}(t)
\chi^+_{-\vec k}(t')\rangle - \langle \chi^+_{\vec k}(t) \chi^-_{-\vec
k}(t')\rangle \right]  \; .
$$
\noindent  Introducing the Fourier transforms
$$
\delta_{\vec k}(t) = \int \frac{dq_0}{2\pi}~ \delta(q_0, {\vec k})~
e^{-iq_0 t} ~~; ~~ J_{\chi,\vec k}(t) = \int \frac{dq_0}{2\pi}~ J(q_0,
{\vec k}) ~ e^{-iq_0 t}   
$$
\noindent we find
$$
\delta( {\vec k},q_0) = - J_{\chi}( {\vec k},q_0) \;
\rho_{\chi}(k,q_0)   
$$
We now use the tadpole method \cite{noneq} to obtain the equation of
motion for the expectation value of the auxiliary field in leading
order in the large $ N $ limit, thereby  
obtaining an explicit expression for $\rho_{\chi}(q_0,\vec k)$ to this
order. The implementation of the tadpole method begins by shifting 
the auxiliary field 
$$
\chi(\vec x,t) = \delta(\vec x,t) + \tilde{\chi}(\vec x,t) ~~; ~~
\langle \tilde{\chi}(\vec x,t) \rangle =0 
$$
\noindent and requiring that $\langle \tilde{\chi}(\vec x,t) \rangle
=0$ to all orders in perturbation theory. A counterterm is added to 
the Lagrangian to cancel the tadpole contributions so as to make the
expectation value of the auxiliary field to vanish in the absence of 
the source term thus allowing to extract the spectral density
straightforwardly. 

To leading order in the large $ N $ limit we obtain the equation of
motion (after the cancellation of the tadpole term) to be given by 
$$
 \delta(\vec x,t) + \int d^3 x' \;  dt' \;  \Pi_r(\vec x-\vec
 x',t-t') \; \delta(\vec x',t) = -J_{\chi}(\vec x,t)  
$$
\noindent with the retarded polarization given by 
\bea
\Pi_r(\vec x-\vec x',t-t') & = & 2i \frac{\lambda}{N}\sum_{a,b} \left[
\langle \Phi^+_a(\vec x,t)\Phi^+_b({\vec x}',t)  \rangle 
\langle \Phi^+_a(\vec x,t)\Phi^+_b({\vec x}',t)\rangle - \right. \nonumber \\
&& \left. \langle \Phi^+_a(\vec x,t)\Phi^-_b({\vec x}',t)  \rangle 
\langle \Phi^+_a(\vec x,t)\Phi^-_b({\vec x}',t)\rangle \right]\nonumber 
\eea 
In terms of the spatial Fourier transform the equation of motion becomes
$$
\delta_{\vec k}(t)+ \int_{-\infty}^{\infty}dt' \;  \Pi_{k,r}(t-t') \;
\delta_{\vec k}(t') = - J_{\chi,\vec k}(t) 
$$
\noindent and the retarded polarization kernel simplifies to 
\bea
\Pi_{k,r}(t-t')& = & 2i\lambda \int \dbarq \left[(-iG^>_{\vec
q}(t-t'))(-iG^>_{\vec q+\vec k}(t-t'))-(-iG^<_{\vec
q}(t-t'))(-iG^<_{\vec q+\vec k}(t-t'))\right]\Theta(t-t') \nonumber \\ 
&= & 4\lambda \int \dbarq \frac{1}{4 q \; |\vec
k+\vec q|}\left\{(1 + n_{\vec q} + n_{\vec q+\vec
k})\sin\left[(q+|\vec k+\vec q|)(t-t')\right]
\right. \nonumber \\ 
&&  \left.  + (n_{\vec q} - n_{\vec q+\vec k}) \sin\left[(|\vec
k+\vec q|-q)(t-t')\right]   \right\}\Theta(t-t') \; ,\nonumber
\eea 
\noindent using  the representation of the theta function given by
(\ref{thetafunc}) we find the time-Fourier representation of the
retarded polarization to be given by
$$
\Pi_{k,r}(t-t') = \int \frac{dq_0}{2\pi} \;\Pi(q_0,q) \;e^{-iq_0(t-t')} 
$$
The Fourier transform of the polarization is now written as a dispersion
integral in terms of the spectral density as 
$$
\Pi( q_0,q)= -\frac{1}{\pi} \int d\omega \;
\frac{\Pi_I(q,\omega)}{q_0-\omega+i\epsilon} 
$$
where
\bea
\Pi_I(q,\omega)= {2\lambda\pi} \int \dbarp 
\frac{1}{4p|\vec p+ \vec q|} && \left\{
\left[1+n_{\vec q + \vec p}+n_{\vec p}\right]\left[ 
\delta(\omega -|\vec p+ \vec q|-p )
-\delta(\omega +|\vec p+ \vec q|+p
)\right]\right. \nonumber \\
&&\left. +\left[n_{\vec p}-n_{\vec q + \vec p} \right]
\left[ \delta(\omega -|\vec p+ \vec q|+p )
-\delta(\omega +|\vec p+ \vec q|-p )\right]
\right\} \label{specdensN} 
\eea 

Finally, in terms of the time Fourier transform the equation of motion
for the expectation value of the auxiliary field is given by 
$$
\delta(q_0,\vec q)\left[1+\Pi(q_0,q)\right] = - J_{\chi}(q_0,\vec q) 
$$
\noindent and we can read off the propagators for the auxiliary field
in Fourier space
\be
G_{\chi}(q_0,q) = \frac{1}{1+\Pi(q_0,q)}
\label{finspecaux} 
\ee
The series of diagrams that are being summed leading to the propagator
for the auxiliary field is shown in fig. (2).  

We now have all of the elements necessary to obtain $ \rho^>_{\phi^2
\phi^2}(q_0,q), \; \rho^<_{\phi^2 \phi^2}(q_0,q) $, writing
$ \Pi (q_0,q)= \Pi_R(q_0,q) + i  \; \Pi_I( q_0,q) $ and comparing
the imaginary parts of(\ref{rhosrelation}) and (\ref{finspecaux}) and
using the KMS condition (\ref{KMS2}) we finally find  
\bea
&& \lambda \; \rho^>_{\phi^2 \phi^2}(q_0,q) = \frac{1}{\pi}\;
\frac{\Pi_I(q_0,q)\;[1+n(q_0)]}{\left[1+\Pi_R(
q_0,q)\right]^2+\Pi^2_I(q_0,q)} \label{rhopi2great} \\ 
&& \lambda  \; \rho^<_{\phi^2 \phi^2}(q_0,q) = \frac{1}{\pi}\;
\frac{\Pi_I(q_0,q)\;n(q_0)}{\left[1+\Pi_R(q_0,q)\right]^2+\Pi^2_I(q_0,q)}
\label{rhopi2less}\\  
&& n(q_0)= \frac{1}{e^{\beta q_0}-1} \nonumber
\eea
We postpone the evaluation of $ \Pi(q_0,q) $ to Appendix B and now
focus on obtaining the resummed self energy for the order parameter. 

\subsection{Equation of motion for the order parameter}
We now obtain the equation of motion for the order parameter to ${\cal
O}(1/N)$ in the linearized approximation again via the tadpole method
and recognize the self-energy to this order. To this effect we write
the field as in (\ref{fieldsplit}) with
$$
\langle \Phi^i(\vec x,t) \rangle = \varphi(\vec x,t)\; \delta_{i,1} \quad
; \quad \langle \eta^i(\vec x,t)\rangle =0  
$$
\noindent where we chose the particular  direction ``1'' by choosing
explicitly the  external source in (\ref{lagrangian}) as $ J^i (\vec
x,t)= J(\vec x,t)\;  \delta^{i,1} $ to  give the 
field an expectation value solely in this direction.  The equation of
motion for $ \varphi(\vec x,t) $ is obtained by imposing that  
$\langle\eta^i(\vec x,t)\rangle =0 $ consistently in the perturbative
expansion. In terms of the spatial Fourier transform of 
the order parameter $ \varphi(t) $ we find
$$
\ddot{\varphi}_k(t)+[k^2 +\delta M^2(T)+M^2_{tad}(T)] \; \varphi_k(t) +
\int_{-\infty}^{\infty} \Sigma_{ret,k}(t-t') \; \varphi_k(t') \;  dt' = J_k(t)
$$
\noindent where $J_k(t)$ is the external source that generates the
initial value problem and $M^2_{tad}(T)\sim \langle (\vec{\Phi}(\vec
x,t)^2\rangle\sim {\cal O}(N) $ is the tadpole contribution which is
the leading order in the large $ N $ limit. The ${\cal O}(1/N)$
contribution to  the self-energy is calculated in terms of the
auxiliary field and is given by 
$$
\Sigma_{ret,k}(t-t') = -\frac{4i\lambda}{N} \int \dbarq
\left[(-iG^{++}_{\vec k+\vec q}(t-t'))\langle \chi^+_{\vec
q}(t)\chi^+_{-\vec q}(t')\rangle -  (-iG^{+-}_{\vec k+\vec
q}(t-t'))\langle \chi^+_{\vec q}(t)\chi^-_{-\vec q}(t')\rangle \right] 
$$
\noindent with $\langle \chi^+_{\vec q}(t)\chi^+_{-\vec q}(t')\rangle$
and  $\langle \chi^+_{\vec q}(t)\chi^-_{-\vec q}(t')\rangle$ 
the {\em full} propagators up to ${\cal O}(1/N)$ given by
(\ref{PPrelation}-\ref{PMrelation}) in terms of the spectral
representations given by (\ref{rhogreatphi2}-\ref{rhopmphi2}) with the spectral
densities given in terms of the self-energy of the auxiliary field 
by (\ref{rhopi2great}-\ref{rhopi2less}).   The contribution to the
propagator of the scalar field up to order  
${\cal O}(1/N)$ is 
depicted in fig. 4. The contribution to the auxiliary field
propagators from the delta functions $\pm i\delta(t-t')$ gives a local
tadpole which is cancelled along with the leading order ${\cal O}(1)$
tadpole contribution by the counterterm to set the theory 
at the critical point up to this order in the large $ N $
expansion. Using the spectral representation for the propagators of
the auxiliary  
field and the free field propagators for the bosonic fields given by
(\ref{ggreat})-(\ref{gsmall}) and after some straightforward algebra 
using the relation $1+n(-q_0) = -n(q_0)$ we finally obtain 
$$
\Sigma_{ret,k}(t-t') = \int_{-\infty}^{\infty} \tilde{\rho}(\omega, k)
\sin\left[\omega(t-t')\right] \;  d\omega 
$$
with $ \tilde{\rho}(\omega,k) $ given by
eqs.(\ref{tilderholargeN})-(\ref{realpi}). 

\section{The retarded polarization of the auxiliary field}
The spectral density (\ref{specdensN}) is the same as
(\ref{specdensfin}) up to the factor $(N+2)/N$, a relatively
straightforward calculation with the Bose-Einstein distribution
functions for massless particles then leads to  
\be
\Pi_I(q_0,q)= \frac{\lambda}{8\pi} \left\{ \Theta(|q_0|-q) \;
\mbox{sign}(q_0)+ \frac{2T}{q}
\ln\left[\frac{1-e^{-\frac{|q_0+q|}{2T}}}{1-e^{-\frac{|q_0-q|}{2T}}}\right]\right\}
\label{pimN} 
\ee
The real part must be obtained via the dispersive integral
(\ref{realpi}). We are only interested in the finite temperature
contribution to both the real and imaginary part therefore we only
consider the second term in (\ref{pimN}). It proves convenient to
write the polarization as a dispersion relation 
$$
\Pi(q_0-i0,q) = \frac{1}{\pi} \int_{-\infty}^{+\infty} d\omega \;
\frac{\Pi_I(\omega,q)}{\omega-q_0+i0} 
$$
\noindent and to analytically continue $ 0 + i q_0 =s$. Using the
fact that $ \Pi_I(\omega,q) $ is an odd function of $ \omega $ we obtain
the dispersion relation
\be\label{intjod}
\Pi(s,q) = \frac{\lambda T}{2\pi^2 q} \int^{\infty}_0 d\omega \; 
\frac{\omega }{\omega^2 +s^2}~  
\ln\left[\frac{1-e^{-\frac{|\omega+q|}{2T}}}{1-e^{-\frac{|\omega-q|}{2T}}}\right]
- \frac{\lambda}{8\pi^2}\ln\left[{q^2 + s^2 \over \mu^2}\right]
\; .
\ee
where $ \mu^2 $ is a subtraction point.

We compute this integral using the sine-Fourier transform as follows. The integrand of eq.(\ref{intjod}) is the product of
two odd functions of $\omega$ :
$$
f_1(\omega) = \frac{\omega }{\omega^2 +s^2} \quad \mbox{and} \quad
f_2(\omega)
=\ln\left[
\frac{1-e^{-\frac{|\omega+q|}{2T}}}{1-e^{-\frac{|\omega-q|}{2T}}}\right]
\; .
$$
We can then apply the Plancherel formula 
$$
\int_0^{\infty} d\omega \; f_1(\omega) \; f_2(\omega) = 
\int_0^{\infty} dx \; {\tilde f}_1(x) \; {\tilde f}_2(x)
$$
where $ {\tilde f}_1(x) $ and $ {\tilde f}_2(x) $ are the sine-Fourier
transforms of  $ f_1(\omega) $ and $ f_2(\omega) $,
respectively. That is,
$$
{\tilde f}_i(x) = \sqrt{2 \over \pi} \int_0^{\infty} d\omega \; 
f_i(\omega) \sin \omega x
$$
where $ i = 1, 2 $. We find \cite{grad}
$$
{\tilde f}_1(x) = \sqrt{\pi \over 2}  \; e^{-sx}
$$
and
$$
{\tilde f}_2(x) = \sqrt{2 \over \pi} \; { \sin q x \over 2 \; T \; x^2 }
\left[2\pi T \; x \; \coth \left(2\pi T \, x \right)-1\right] \; .
$$
We have now that 
$$
\Pi(s,q) = {\lambda  \over (2\pi)^2 \; q } \int_0^{\infty} dx \; {
e^{-sx}\over x^2} \; \sin q x \; \left[2\pi T \, x \; \coth\left( 2\pi T
\, x\right)  -1\right] - \frac{\lambda}{8\pi^2}\ln{q^2 +
s^2 \over \mu^2} \; .
$$
It is convenient to split this integral into two terms,
\bea
\Pi(s,q) &=& {\lambda  \over (2\pi)^2 \, q} \left\{\int_0^{\infty} dx
\; {e^{-sx}\over x} \left[ 1 - { \sin q x \over x} \right] \right.\cr \cr
&+& \left. \int_0^{\infty} dx \; {e^{-sx}\over x}\left[2\pi T  \;
\coth\left( 2\pi T \, x\right)\, \sin q x -1\right]\right\}-
\frac{\lambda}{8\pi^2}\ln{q^2 + s^2 \over \mu^2} \; .\nonumber
\eea
We carried out the integration explicitly with the result\cite{grad}
$$
\Pi(s,q) = \frac{\lambda }{(2\pi)^2}\left\{-1+
\ln\frac{\mu}{4\pi T}+\frac{s}{q} \; \mbox{arctg}{q \over s}\;
-\frac{i\pi T}{q} \ln\left[\frac{\Gamma\left( \frac{is+q}{4\pi
iT}\right)\Gamma\left(1+\frac{is+q}{4\pi iT} \right)}{\Gamma\left(
\frac{is-q}{4\pi iT}\right)\Gamma\left(1+\frac{is-q}{4\pi iT} \right)} \right]
\right\}
$$
where we used Malmsten formula for the Gamma functions\cite{grad}. 

Back in real frequencies we have,
$$
\Pi(q_0 \pm i0,q) = \Pi_R(q_0,q) \pm i \, \Pi_I(q_0,q)
$$
where $ \Pi_I(q_0,q) $ is given by eq.(\ref{pimN}) and
\bea
\Pi_R(q_0,q) &=& \frac{\lambda }{(2\pi)^2}\left\{ {\pi^2 T \over q}
\left[ \theta(q-q_0) - \theta(-q-q_0) \right] + \ln\frac{\mu}{4\pi T}
\right. \cr \cr
&+& \left. { q_0 \over 2 \, q} \ln\left|{q+q_0\over q-q_0}\right| + 
\frac{2\pi T}{q} \mbox{Im} \ln \left[ \Gamma\left(1+\frac{q-q_0}{4\pi
iT} \right) \Gamma\left(1+\frac{q+q_0}{4\pi iT} \right) \right]
\right\}\; . \nonumber
\eea
The limit $T/q   \gg  >1$ can be taken in a straightforward manner 
and we obtain the high temperature limit of the polarization to be given by
\bea
\Pi(q_0,q+i\epsilon) &= &\frac{i\; \lambda \, T}{4\pi  q} 
\ln\left[\frac{q_0+i\epsilon + q}{q_0+i\epsilon -q}\right] \nonumber \\
&+&\frac{\lambda}{(2\pi)^2} \left[ \ln\frac{\mu}{4\pi T} + { q_0 \over 2 q}
\ln\left(\frac{q_0+i\epsilon + q}{q_0+i\epsilon -q}\right) + 2 \gamma \right]+
{\cal O}\left({1\over T}\right)\nonumber
\eea
where $ \gamma $ is the Euler-Mascheroni constant.
%%%%%%%%bibliography goes here
 
%%%%%%%%%%%%%begin figures%%%%%%%%%%%%%%
%%%%%%%%%%%%%%%figure 1%%%%%%%%%%%
%%%%%%%%%%%function F[x] that enters in the two-loops damping rate 
\begin{figure}[ht]
\epsfig{file=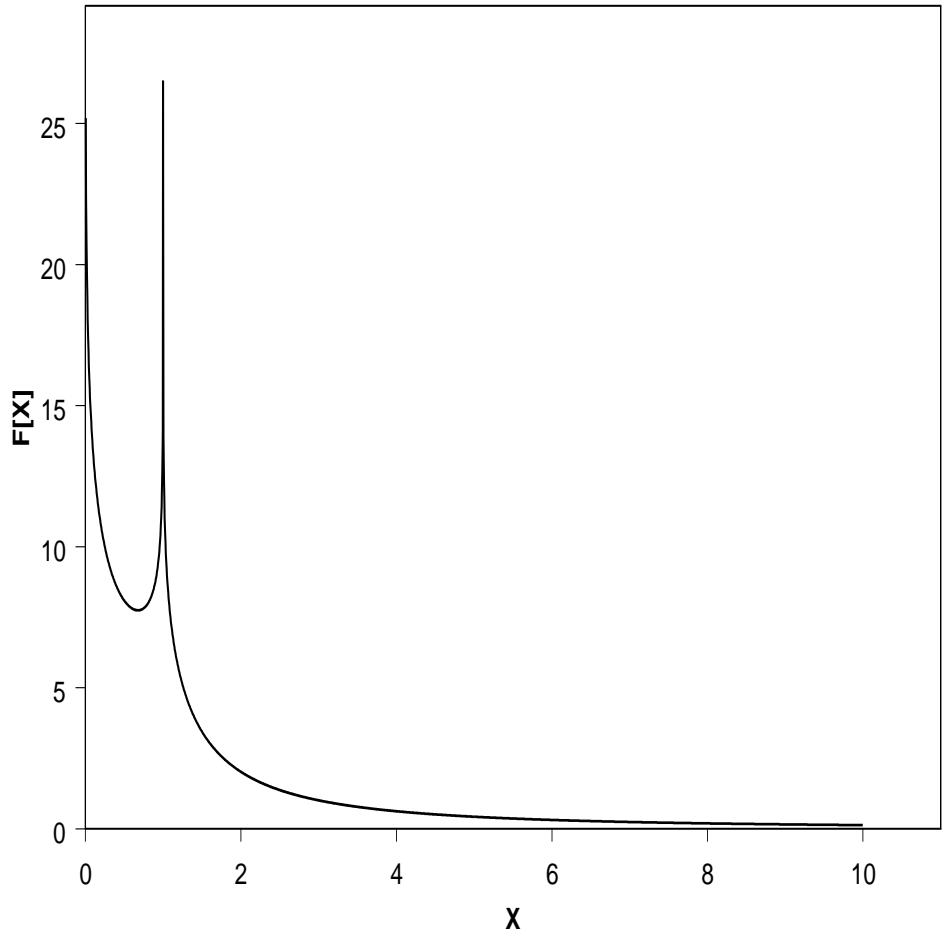,width=6in,height=6in}
\vspace{.1in}
\caption{ The function $F[x]$ vs. $x=q/k$.
\label{fig1}}
\end{figure}  
%%%%%%%%%%%end figure 1%%%%%%%%%%%%
%%%%%%%%%figure 2 prop for aux. field
%\begin{center}
\begin{figure}[ht]
\epsfig{file=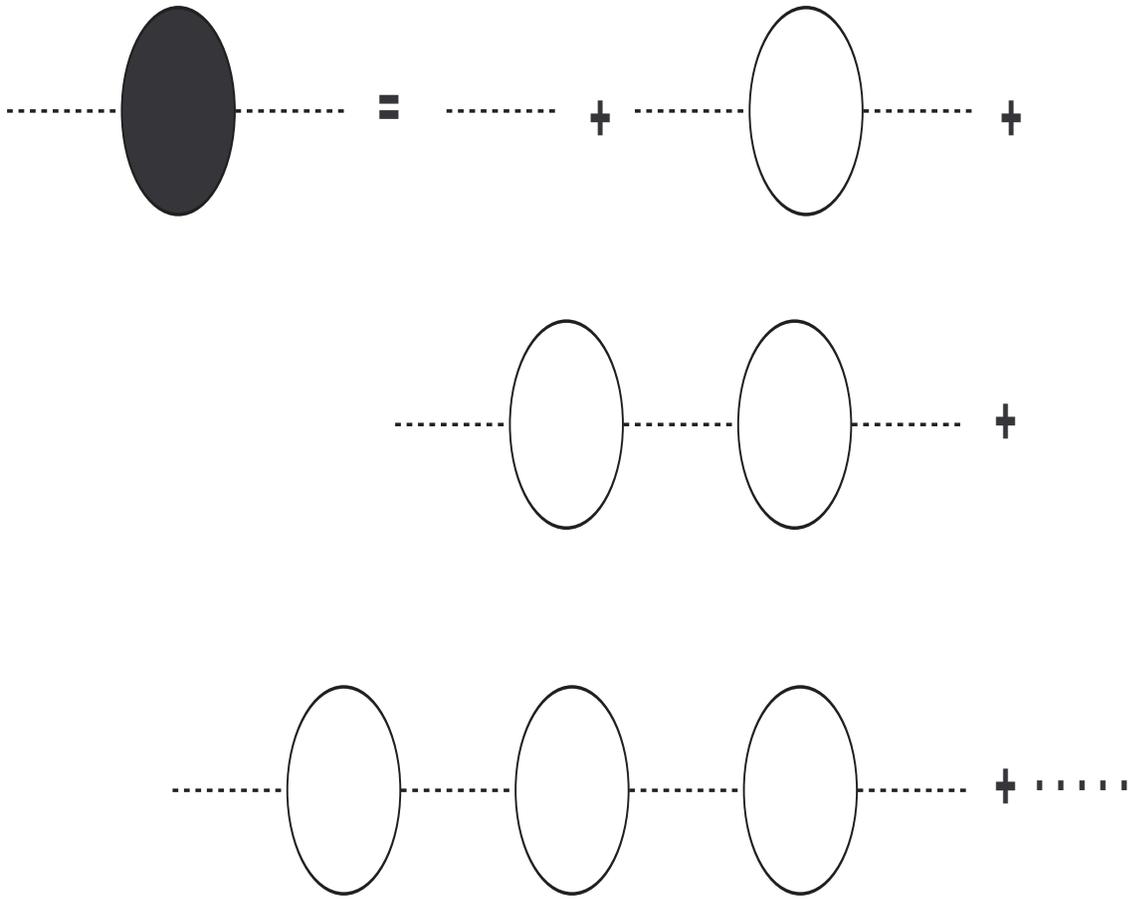,width=7in,height=5in}
\vspace{.1in}
\caption{ Propagator for the auxiliary field in leading order in the large $ N $ limit. There is a factor $\sqrt{\lambda \over N}$ for each vertex and a factor $N$ for each bubble. The propagator is of ${\cal O}(1)$ in the large $ N $ limit. 
\label{fig2}}
\end{figure}  
%%%%%%end figure 2 %%%%%%%%%
%%%%%%%%%figure 3 scatt amplitude in large N
\begin{figure}[ht]
\epsfig{file=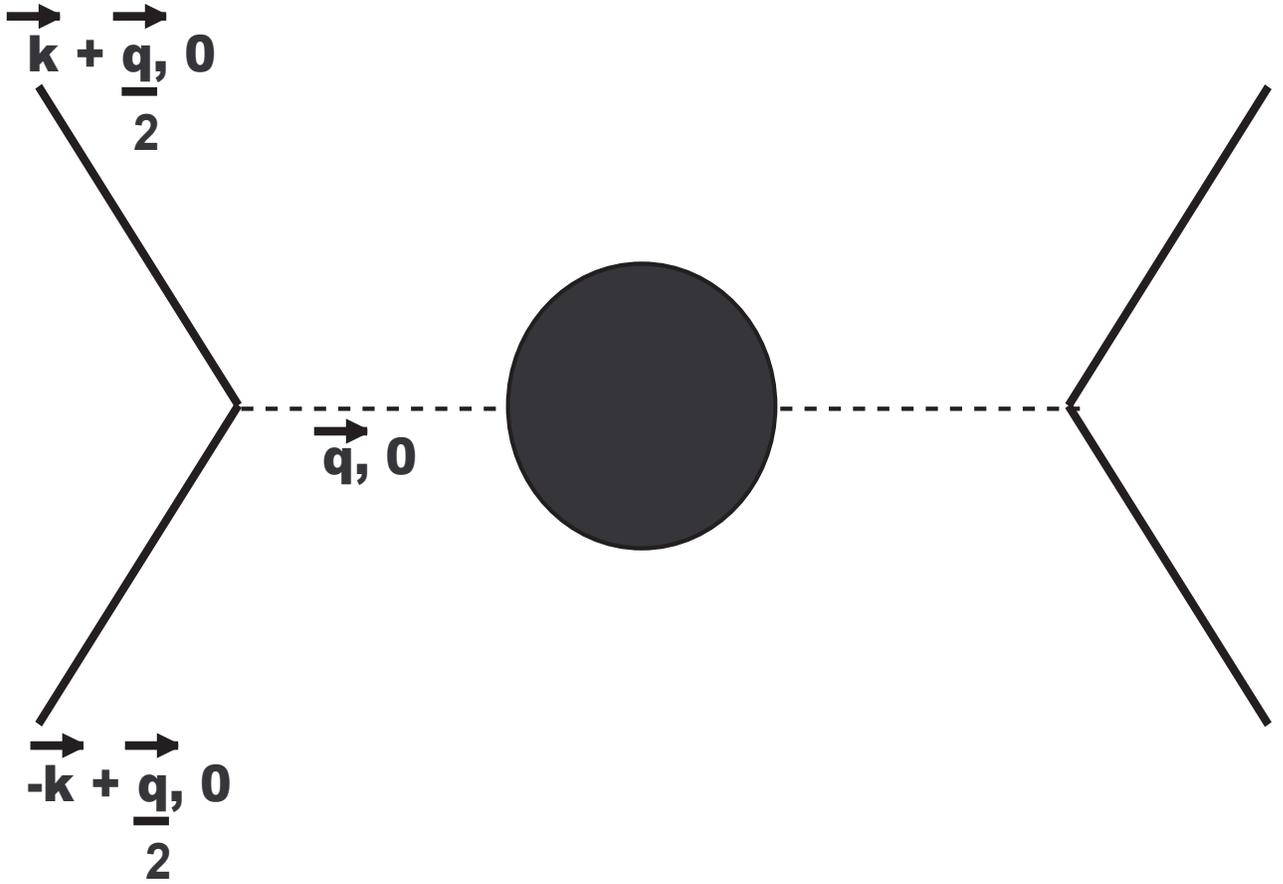,width=7in,height=5in}
\vspace{.1in}
\caption{Two particle s-channel scattering amplitude in the static
and large $ N $ limit. $\vec q,~ q_0=0$ are the transferred momentum
and frequency carried by the propagator of the auxiliary field.  
\label{fig3}}
\end{figure}  
%%%%%%end figure 3 %%%%%%%%%
%%%%%%%%%figure 4  large $ N $ self energy
\begin{figure}[ht]
\epsfig{file=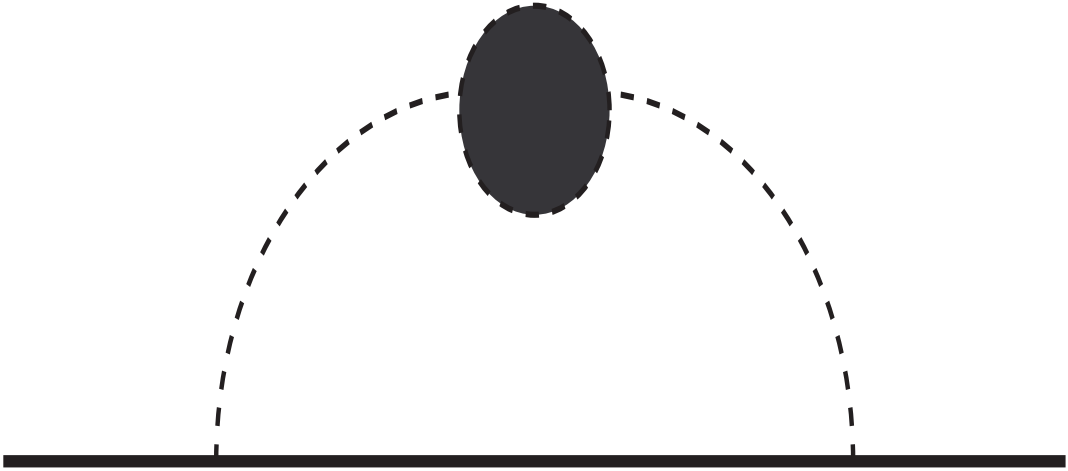,width=6in,height=3in}
\vspace{.1in}
\caption{ The self-energy of the scalar field at order $1/N$. 
\label{fig4}}
\end{figure}  
\begin{figure}[ht]
%\begin{turn}{-90}
\epsfig{file=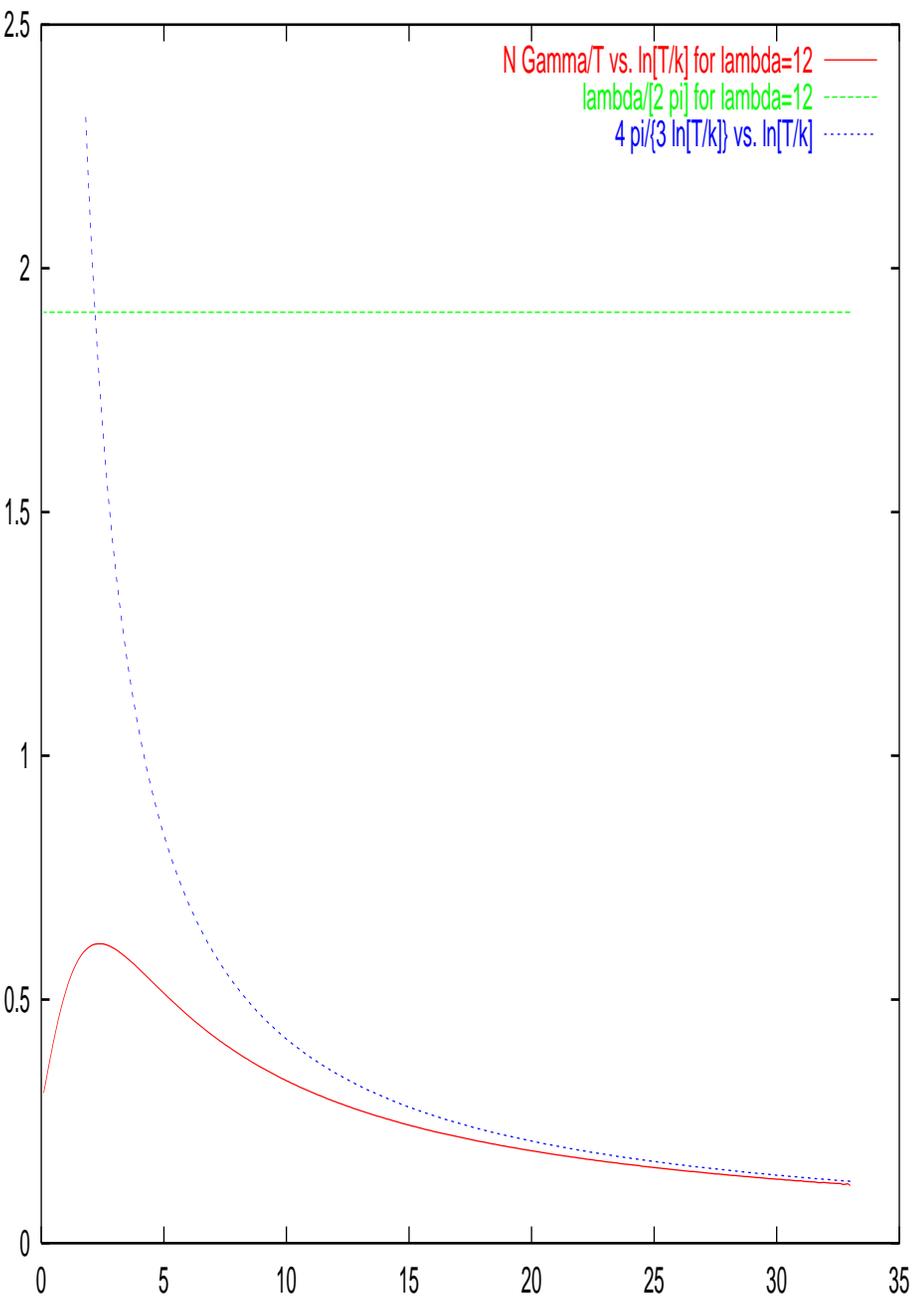,width=7in,height=5in}
%\end{turn}
\vspace{.1in}
\caption{$ N \; \Gamma(k,T)/T $ vs. $ \ln[T/k] $ for $ \lambda  = 12.0  $. We
also plot the classical value $ \lambda/[2 \pi] $ and the asymptotic
ultrasoft behavior $  N \; \Gamma(k,T)/T = {4 \pi \over 3 \ln[T/k]}
$. We see in this strong coupling regime that the ultrasoft
asymptotics correctly describes the relaxation rate. \label{fig5}}
\end{figure}  

\begin{figure}[ht]
%\begin{turn}{-90}
\epsfig{file=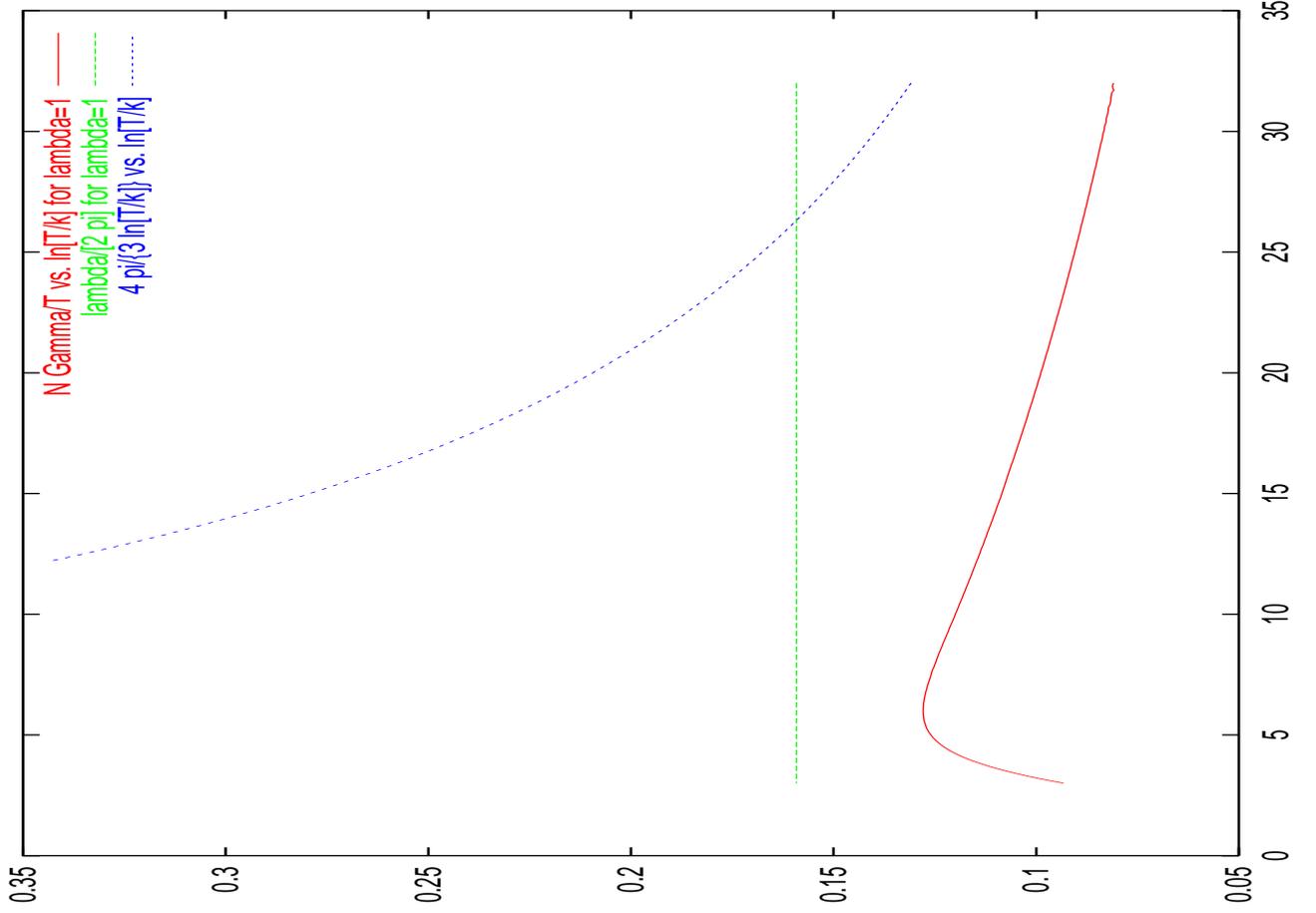,width=7in,height=5in}
%\end{turn}
\vspace{.1in}
\caption{$ N \Gamma(k,T)/T $ vs. $ \ln[T/k] $ for $ \lambda  = 1.0  $. We
also plot the classical value $ \lambda/[2 \pi] $ and the asymptotic
ultrasoft behavior $  N \Gamma(k,T)/T = {4 \pi \over 3 \ln[T/k]} $.
For this intermediate coupling regime the classical approximation provides 
a qualitative a estimate whereas the ultrasoft regime will be reached 
for  $ k/T \ll 3.\; 10^{-7} $.
\label{fig5b}}
\end{figure} 

\begin{figure}[ht]
%\begin{turn}{-90}
\epsfig{file=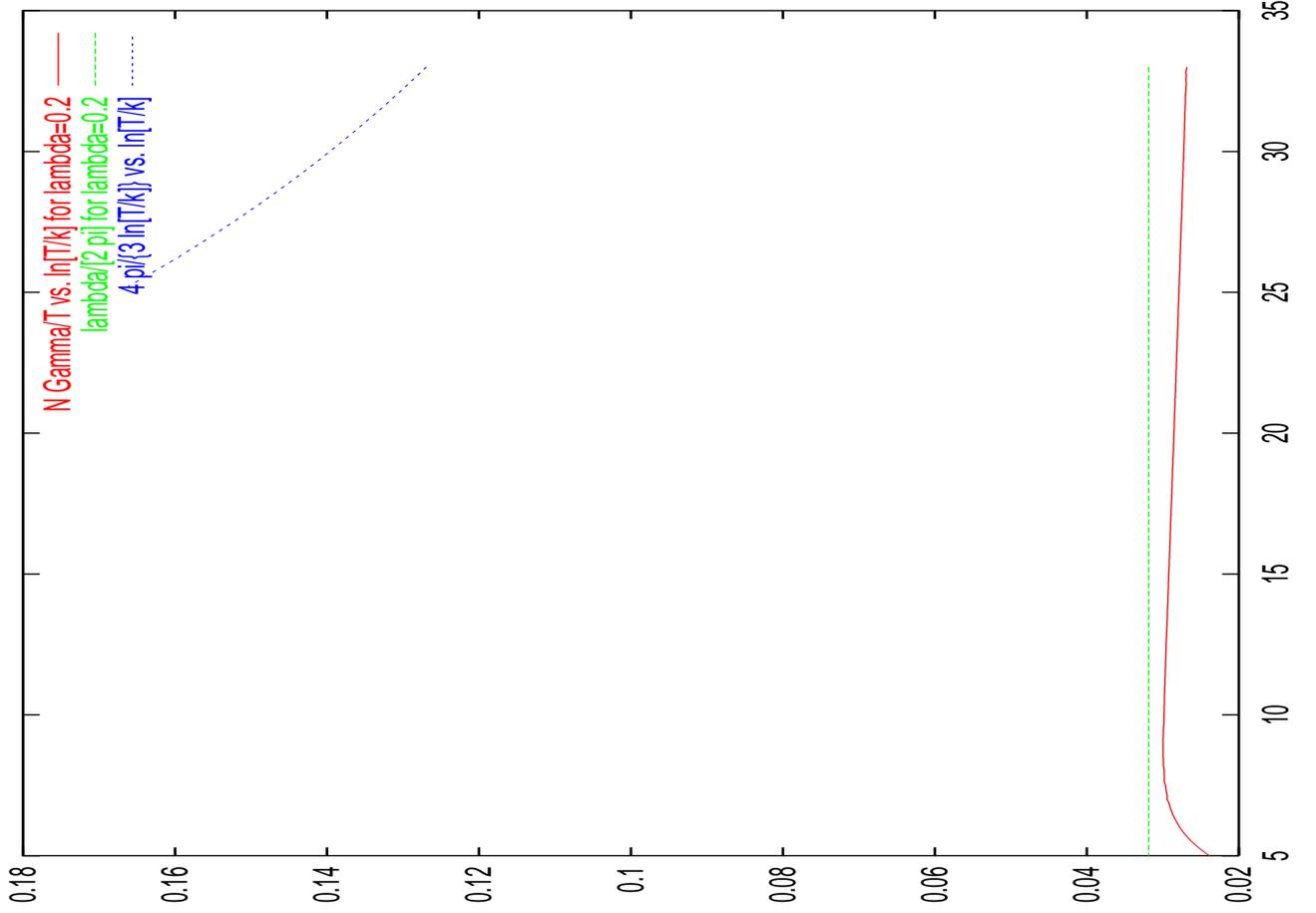,width=7in,height=5in}
%\end{turn}
\vspace{.1in}
\caption{$ N \; \Gamma(k,T)/T $ vs. $ \ln[T/k] $ for $ \lambda  = 0.2 $. We
also plot the classical value $ \lambda/[2 \pi] $ and the asymptotic
ultrasoft behavior $  N\; \Gamma(k,T)/T = {4 \pi \over 3 \ln[T/k]} $. In
the small coupling regime the classical approximation  describes very well
the behaviour of the relaxation rate.The ultrasoft regime will only be
reached for extremely small momenta $ k/T \ll  8. \; 10^{-30} $
\label{fig5c}}
\end{figure} 

\begin{figure}[ht] 
\begin{center}
\epsfig{file=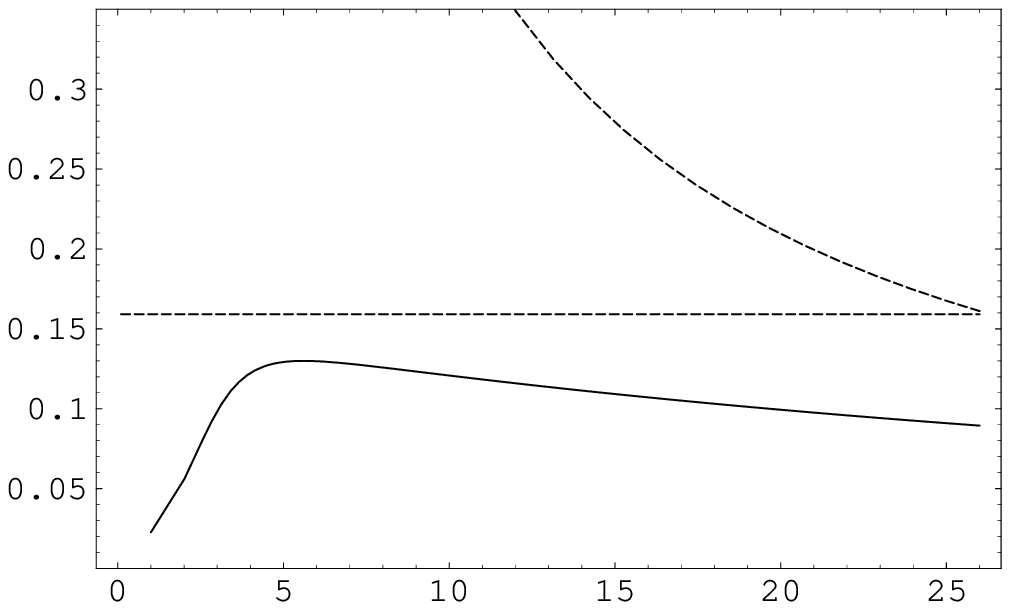,width=7in,height=5in}
  \end{center}
  \caption{{\small Damping rate $ N \; \Gamma_0(m_T,T)/T $ vs. $ \ln[T/m_T] $ 
   for homogenous configurations  ($k=0,\;
   T\neq T_c$) compared with the classical approximation
  $ N \; \Gamma_{cl}(m_T,T)/T =\lambda/(2\pi) $ and the asymptotic
   expression in the medium
coupling regime $\lambda= 1.0 $.}}
\end{figure}

\begin{figure}[ht] 
  \begin{center}
\epsfig{file=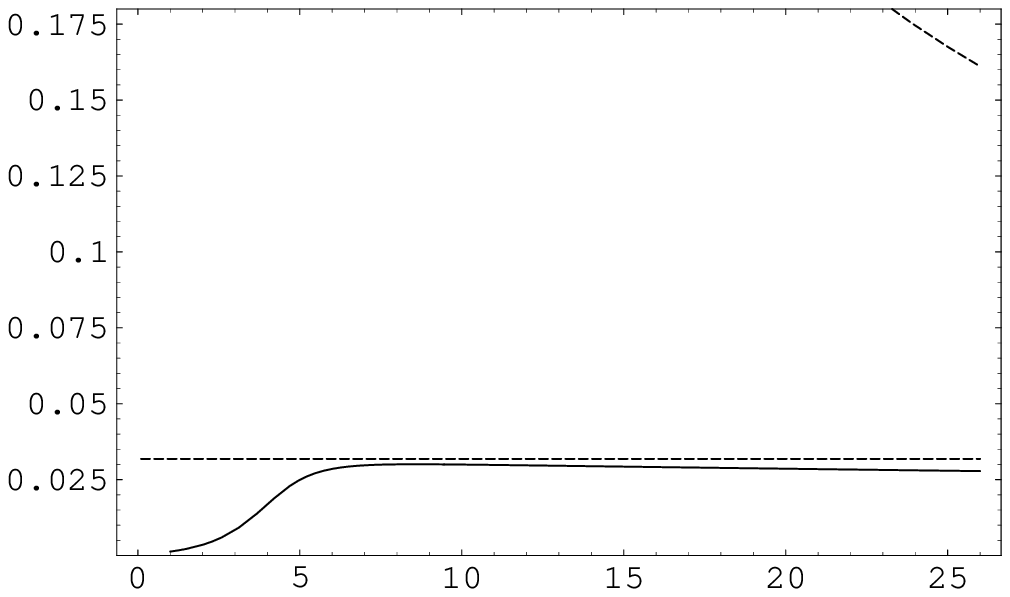,width=7in,height=5in}
  \end{center}
  \caption{{\small Damping rate $ N \; \Gamma_0(m_T,T)/T $ vs. $ \ln[T/m_T] $ 
  for homogenous configurations ($k=0,\;
  T\neq T_c$) compared with the classical approximation
  $ N \; \Gamma_{cl}(m_T,T)/T =\lambda/(2\pi) $  and the asymptotic
   expression in the small
coupling regime $\lambda=0.2$.}}
\end{figure}
%%%%%%%%%figure 7 <\chi^+ \chi^+> and <\Phi^+ \Phi^+> diagrams
%\begin{center}
\begin{figure}[ht]
\epsfig{file=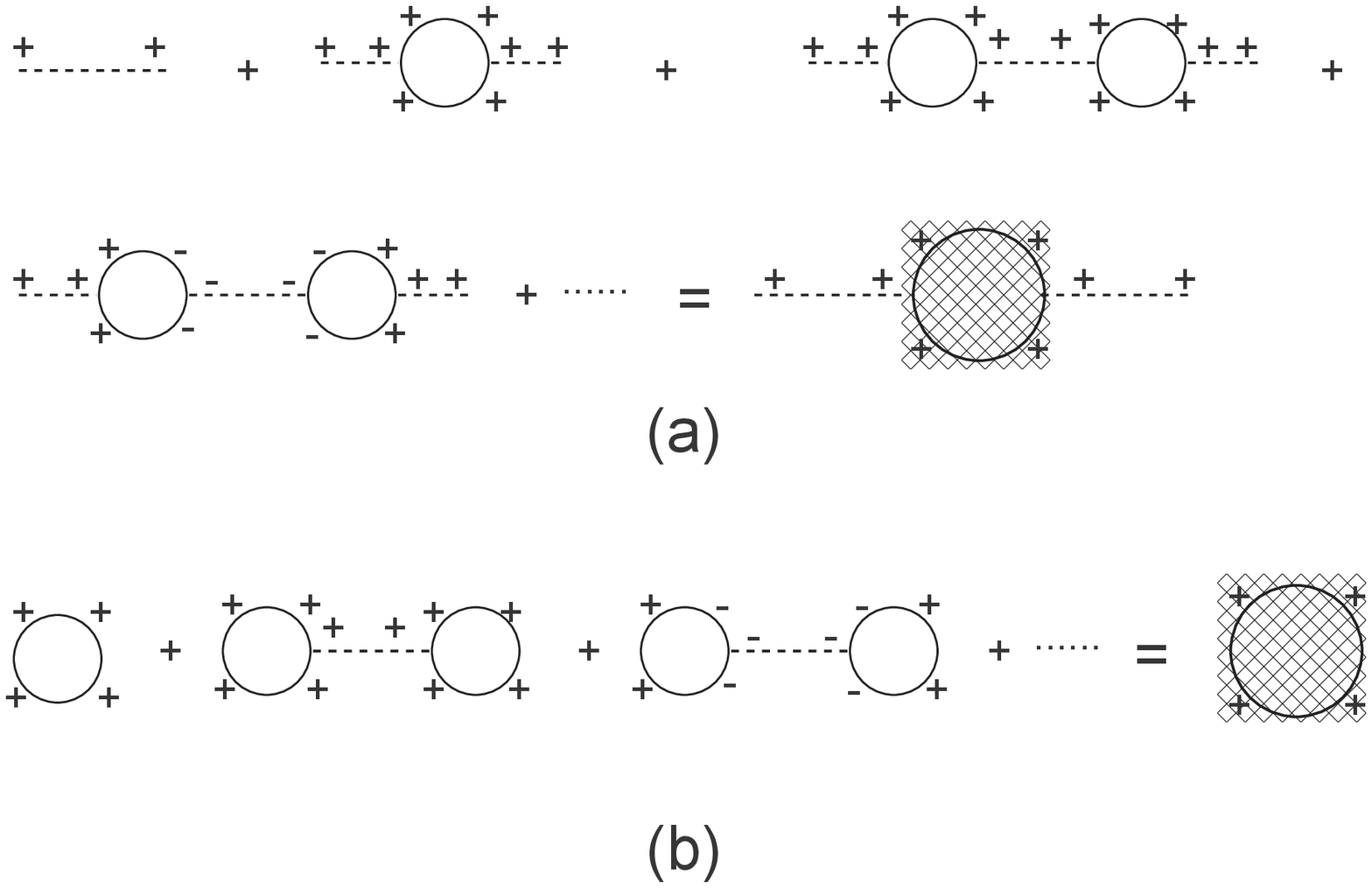,width=7in,height=5in}
\vspace{.1in}
\caption{(a): Feynman diagrams for the correlator $\langle \chi^+(\vec
x,t) \chi^+ (\vec x',t')\rangle$, (b):Feynman diagrams for the
correlator $\langle (\vec \Phi^+(\vec x,t))^2 (\vec \Phi^+ (\vec
x',t'))^2 \rangle$.\label{fig7}}
\end{figure}  
%\end{center}
%%%%%%end figure 7 %%%%%%%%%
%%%%%%%%%figure 8 <\chi^+ \chi^-> and <\Phi^+ \Phi^-> diagrams
\begin{figure}[ht]
\epsfig{file=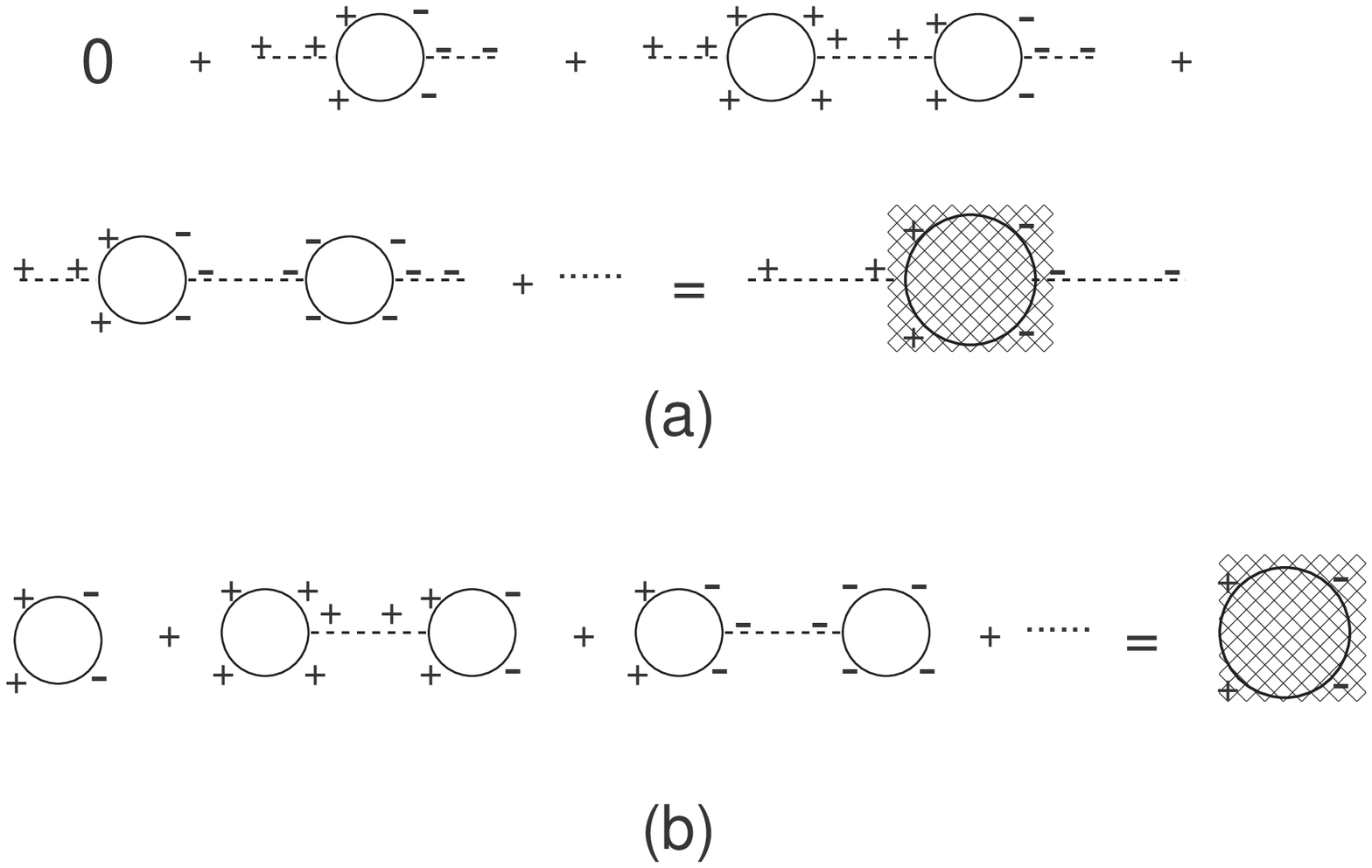,width=7in,height=5in}
\vspace{.1in}
\caption{ (a): Feynman diagrams for the correlator $\langle
\chi^+(\vec x,t) \chi^- (\vec x',t')\rangle$, (b): 
Feynman diagrams for the correlator $\langle (\vec \Phi^+(\vec x,t))^2
( \vec \Phi^- (\vec x',t'))^2 \rangle$. 
\label{fig8}}
\end{figure}  
%%%%%%end figure 8 %%%%%%%%%
%%%%%%%%%figure 9 <\chi^- \chi^-> and <\Phi^- \Phi^-> diagrams
\begin{figure}[ht]
\epsfig{file=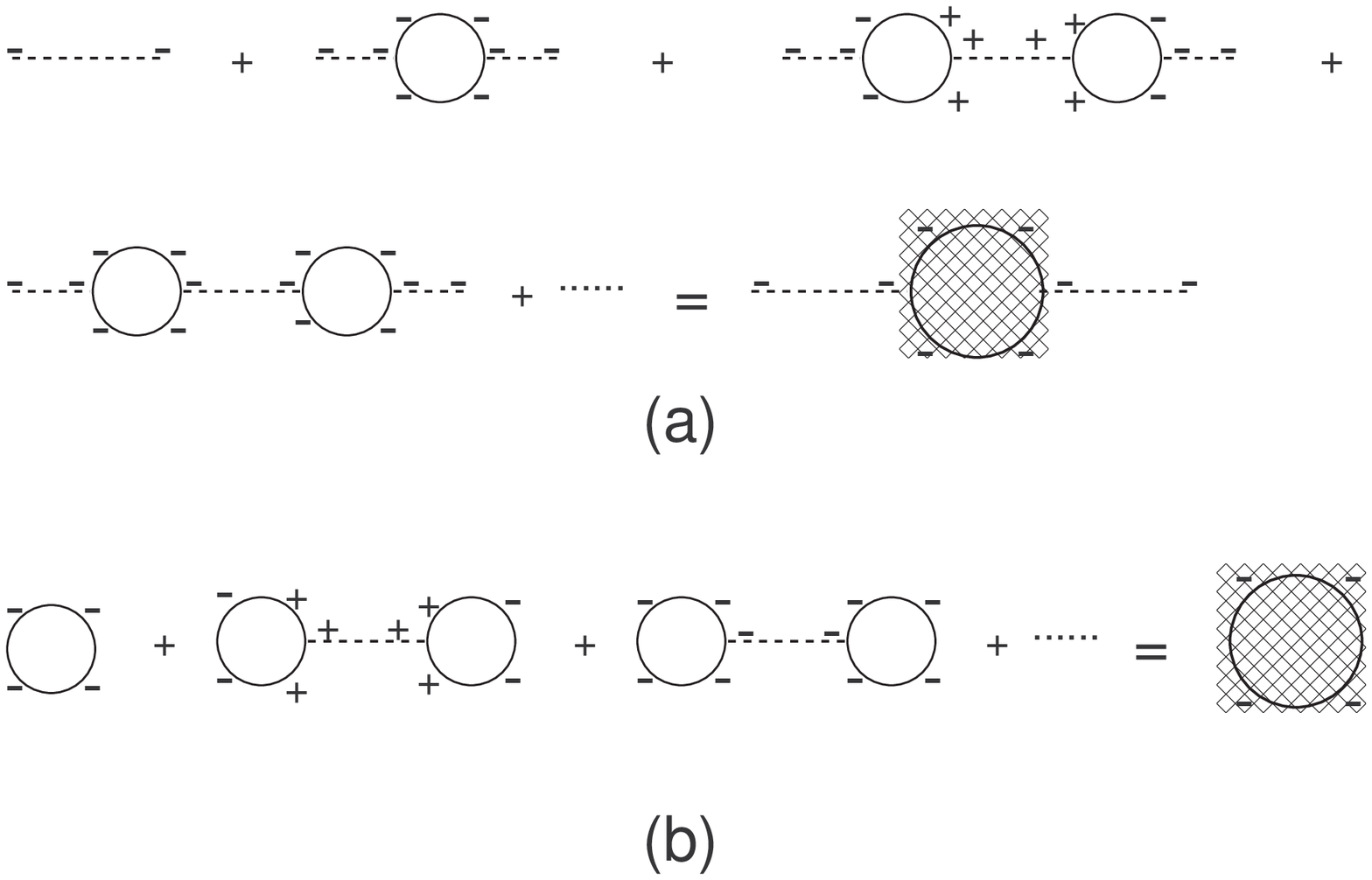,width=7in,height=5in}
\vspace{.1in}
\caption{ (a): Feynman diagrams for the correlator $\langle \chi^-(\vec x,t) \chi^- (\vec x',t')\rangle$,  (b):
Feynman diagrams for the correlator $\langle (\vec \Phi^-(\vec x,t))^2 ( \vec \Phi^- (\vec x',t'))^2 \rangle$.
\label{fig9}}
\end{figure}  
%%%%%%end figure 9 %%%%%%%%%
%%%%%%%%%%%%%%%%%%end figures %%%%%%%%%%%%%%%%%%%%%%%%%%%%%%%%%%%%%%
\end{document}